\documentclass[runningheads]{llncs}
\usepackage{amscd}
\usepackage{amssymb}
\usepackage{amsmath}
\usepackage{stmaryrd}
\usepackage{mathrsfs}
\usepackage{times}
\usepackage{latexsym}
\usepackage{hhline}

\usepackage[latin1]{inputenc} % From LaTeX distribution
\usepackage{calc}         % From LaTeX distribution
\usepackage{graphicx}     % From LaTeX distribution
\usepackage{ifthen}       % From LaTeX distribution
\usepackage{subfigure}    % From CTAN/macros/latex/contrib/supported/subfigure
\usepackage{pst-all}      % From PSTricks
\usepackage{pst-poly}     % From pstricks/contrib/pst-poly
\usepackage{multido}      % From PSTricks

\newenvironment{pf}[1][Proof]{\begin{trivlist}
\item[\hskip \labelsep {\bfseries #1}]}{\end{trivlist}}

\newcommand{\prusa}[0]{Pr{\r{u}}{\v s}a}

\DeclareMathOperator{\LL}{\mathcal{L}}
\DeclareMathOperator{\HH}{\mathcal{H}}
\DeclareMathOperator{\VV}{\mathcal{V}}

\begin{document}

\title{A unifying approach to picture
  grammars\thanks{A preliminary version is \cite{RTG1}. Work partially
  supported by PRIN Project ``Mathematical aspects and emerging
  applications of automata and formal languages'', ESF Programme {\em
    Automata: from Mathematics to Applications (AutoMathA)}, and CNR
  RSTL Project 760 {\em Grammatiche 2D per la descrizione di
    immagini}.}}

\author{Matteo Pradella\inst{1} \and
        Alessandra Cherubini\inst{2} \and
        Stefano {Crespi Reghizzi}\inst{2} 
        }

\institute{CNR IEIIT-MI  \\ 
           \and Politecnico di Milano \\
           P.zza L. da Vinci, 32,
           20133 Milano, Italy\\
           \email{\{alessandra.cherubini, stefano.crespireghizzi,\\
             matteo.pradella\}@polimi.it}}

\maketitle

\begin{abstract}
  Several old and recent classes of picture grammars, that
  variously extend context-free string grammars in two dimensions, are based on
  rules that rewrite arrays of pixels. Such grammars can be unified and extended using a tiling based
  approach, whereby the right part of a rule is formalized by means of a
  finite set of permitted tiles. We focus on a simple type of tiling,
  named \emph{regional}, and define the corresponding regional tile
  grammars. They include both Siromoney's (or Matz's) Kolam grammars
  and their generalization by Pr{\r{u}}{\v s}a, as well as Drewes's grid grammars.
  Regionally defined
  pictures can be recognized with polynomial-time complexity by an
  algorithm extending the CKY one for strings.  Regional tile grammars
  and languages are strictly included into our previous tile grammars and
  languages, and are incomparable with
  Giammarresi-Restivo tiling systems (or Wang systems).
\end{abstract}

\section{Introduction}\label{SectionIntroduction}
Since the early days of formal language theory, considerable research
effort has been spent towards the objective of extending grammar based
approaches from one to two dimensions (2D), i.e., from string
languages to picture languages.  Several approaches have been proposed
(and sometimes re-proposed) in the course of the years, which in
different ways take inspiration from regular expressions and from
Chomsky's string grammars, but, to the best of our knowledge, no
general classification or detailed comparison of picture grammars has
been attempted.  It is fair to say that the immense success of
grammar-based approaches for strings, e.g. in compilation and natural
language processing, is far from being matched by picture
grammars. Several causes for this may exist.  First, the lack of
broadly accepted reference models has caused a dispersion of research
efforts. Second, the algorithmic complexity of parsing algorithm for
2D languages has rarely been considered, and very few efficient
algorithms, and fewer implementations, exist. Last, but not least,
most grammar types have been invented by theoreticians and their
applicability in picture or image processing remains to be seen.
\par
We try to remove, or at least to partially offset, the first two
 causes, thus hoping to set in this way the ground for applied
 research on picture grammars. First, we offer a new simple unifying
 approach encompassing most existing grammar models, based on the
 notion of picture tiling. Then, we introduce a new type of grammar,
 called \emph{regional} that is more expressive than several existing
 types, yet it offers a polynomial-time parsing algorithm.
\par
We outline how several classical models of picture grammars based
on array rewriting rules can be unified by a tiling based approach. A
typical rewriting rule replaces a pixel array, occurring in some
position in the picture, by a right part, which is a pixel array of
equal size.  Each grammar type considers different forms of rewriting
rules, that we show how can be formalized using more or less general
sets of tiles. In particular, we focus on a simple type of tile sets,
those of \emph{regional tile grammars}. This new class generalizes some classical models, yet it is
proved to permit efficient, polynomial-time recognition of pictures by
an approach extending the classical Cocke-Kasami-Younger (CKY)
algorithm \cite{CKY} of context-free (CF) string languages.
\par
From the standpoint of more powerful grammar models, regional tile
grammars correspond to a natural restriction of our previous
\emph{tile} (rewriting) \emph{grammars} (TG) \cite{TRG1,CCPS}.
For such grammars, a rule replaces a rectangular area filled with a
nonterminal symbol with a picture belonging to the language defined by
a specified set of tiles over terminal or nonterminal symbols. It is
known that the TG family dominates the family of languages defined by
the \emph{tiling systems} (TS) of Giammarresi and Restivo
\cite{GR-recognizable-pl} (which are equivalent to  Wang systems
\cite{Wang}\cite{labelled-wang1997}), and that the latter are
NP-complete with respect to picture recognition time complexity.  The
new model enforces the constraint that the local language used to
specify the right part of a rule is made by assembling a finite number of
homogeneous rectangular pictures. Such tiling is related to Simplot's
\cite{Simplot99} interesting closure operation on pictures.
\par
Regional tile grammars are then shown to dominate other grammar
types. The first is the classical Kolam grammar type of Siromoney
\cite{GSiromoney-RSiromoney-KKrithivasan:73b} (which, in its
context-free form, is equivalent to the grammars of Matz
\cite{STACS::Matz1997}); it is less general because the right parts of
grammar rules must be tiled in ways decomposable as vertical and horizontal
concatenations. Three other grammar families are then shown to be less
general: {\em \prusa{}'s type} \cite{Prusa2004}, \emph{grid grammars}
\cite{Drewes2003}, and \emph{context-free matrix grammars}
\cite{GSiromoney-RSiromoney-KKrithivasan:72}.  The language inclusion
properties for all the above families are thus clarified.
\par
The presentation continues in Section \ref{SectionBasicDef} with
preliminary definitions, then in Sections
\ref{SectionTileGrammars} and \ref{SectionRegionalGrammars}
with the definition of tile grammars, their regional variant, and relevant examples.
In Section \ref{SectionParsingRTG} we present the parsing algorithm and
prove its correctness and complexity. In
Section \ref{SectionComparisons} we compare regional tile grammars and
languages with other picture language families.  
The paper concludes by summarizing the main results.

%%% Local Variables: 
%%% mode: latex
%%% TeX-master: "paper"
%%% End: 

\section{Basic definitions}\label{SectionBasicDef}

The following notation and definitions are mostly from \cite{GR-two-dl} and \cite{TRG1}.

\begin{definition} \label{basics} Let $\Sigma$ be a finite alphabet. A
  two-dimensional array of elements of $\Sigma$ is a picture over
  $\Sigma$. The set of all pictures over $\Sigma$ is $\Sigma^{++}$. A
  picture language is a subset of $\Sigma^{++}$.

  For $h,k\geq 1$, $\Sigma^{(h,k)}$ denotes the set of pictures of
  size $(h,k)$ (we will use the notation $|p| = (h,k), |p|_{row} = h,
  |p|_{col} = k$). \# $\notin \Sigma$ is used when needed as a {\em
    boundary symbol}; $\hat{p}$ refers to the bordered version of
  picture $p$. That is, for $p \in \Sigma^{(h,k)}$, it is
\[
p =
\begin{array}{ccc}
p(1,1) & \ldots & p(1,k) \\
\vdots & \ddots & \vdots \\
p(h,1) & \ldots & p(h,k)
\end{array} \;\;\;\;\;
\hat{p} =
\begin{array}{ccccc}
\# & \# & \ldots & \# & \# \\
\# & p(1,1) & \ldots & p(1,k) & \# \\
\vdots & \vdots & \ddots & \vdots & \vdots \\
\# & p(h,1) & \ldots & p(h,k) & \# \\
\# & \# & \ldots & \# & \# \\
\end{array}
\]
A {\em pixel} is an element $p(i,j)$ of $p$. If all pixels are
identical to $C \in \Sigma$ the picture is called $C$-{\em
  homogeneous} or $C$-picture.

{\em Row and column concatenations} are denoted $\ominus$ and $\obar$,
respectively. $p \ominus q$ is defined iff $p$ and $q$ have the same
number of columns; the resulting picture is the vertical juxtaposition
of $p$ over $q$. $p^{k\ominus}$ is the vertical juxtaposition of $k$
copies of $p$; $p^{+\ominus}$ is the corresponding closure. $\obar,
^{k\obar}, ^{+\obar}$ are the column analogous.
\end{definition}

\begin{definition} \label{subpicture}
  Let $p$ be a picture over $\Sigma$.
The {\em domain} of a picture $p$ is the set $\mathrm{dom}(p)=\{1, 2,
\ldots, |p|_{row}\}\times \{1, 2, \ldots, |p|_{col}\}$.
 A {\em subdomain} of $\mathrm{dom}(p)$ is
  a set $d$ of the form $\{x, x+1, \ldots, x'\}\times \{y, y+1, \ldots,y'\}$ where
  $1\leq x\leq x'\leq |p|_{row},\ 1\leq y\leq y'\leq |p|_{col}$. We
  will often denote a subdomain by using its top-left and bottom-right
  coordinates, in the previous case the quadruple $(x, y; x', y')$.

  The set of subdomains of $p$ is denoted $D(p)$. Let $d=\{x,\ldots,
  x'\}\times \{y,\ldots,y'\}\in D(p)$, the subpicture
  $\mathrm{spic}(p,d)$ associated to $d$ is the picture of size
  $(x'-x+1,\ y'-y+1)$ such that $\forall i\in \{1, \ldots, x'-x+1\}$
  and $\forall j \in \{1, \ldots, y'-y+1\}$, $\mathrm{spic}(p,d)(i,j)=
  p(x+i-1,y+j-1)$.

  A subdomain is called $C$-homogeneous (or homogeneous) when
  its associated subpicture is a $C$-picture. $C$ is called the {\em
    label} of the subdomain.

  Two subdomains $d_a = (i_a,j_a; k_a,l_a)$ and $d_b = (i_b,j_b;
  k_b,l_b)$ are {\em horizontally adjacent} (resp. {\em vertically
    adjacent}) iff $j_b = l_a+1$, and $k_b \ge i_a, k_a \ge i_b$
  (resp. $i_b = k_a+1$, and $l_b \ge j_a, l_a \ge j_b$).
  We will call two subdomains adjacent, if they are either vertically or
  horizontally adjacent.

  The {\em translation} of a subdomain $d = (x, y; x', y')$ by
  displacement $(a,b) \in \mathbb{Z}^2$ is the subdomain $d' = (x+a,
  y+b; x'+a, y'+b)$. We will write 
  %$d' = \mathrm{transl}_{(a,b)} d$.
  $d' = d \oplus (a,b)$.
\end{definition}

\begin{definition} \label{homogeneous partition}
  A {\em homogeneous partition} of a picture $p$ is any partition $\pi
  = \{d_1, d_2, \ldots, d_n\}$ of
  $\mathrm{dom}(p)$ into homogeneous subdomains $d_1, d_2, \ldots,
  d_n$. 

  The {\em unit partition} of $p$, written $\mathrm{unit}(p)$, is the
  homogeneous partition of $\mathrm{dom}(p)$ defined by single pixels.

  An homogeneous partition is called {\em strong} if adjacent
  subdomains have different labels.
\end{definition}

We observe that if a picture $p$ admits a strong homogeneous partition
of $\mathrm{dom}(p)$ into subdomains, then the partition is unique and
will be denoted by $\Pi(p)$.
\\
To illustrate, all the pictures in Figure \ref{FigDerivationEx1} but
the last two admit
a strong homogeneous partition, which is depicted by outlining the borders
of the subdomains. The marked partitions of the last two pictures are
homogeneous but not strong, because some adjacent subdomains hold the
same letter.

We now introduce the central concepts of  {\em
  tile}, and {\em local language}.

\begin{definition}\label{local}
  We call {\em tile} a square picture of size (2,2).  We denote by
  $\llbracket p \rrbracket$ the set of all tiles contained in a picture $p$.
\\
  Let $\Sigma$ be a finite alphabet.  A (two-dimensional) language $L
  \subseteq \Sigma^{++}$ is {\em local} if there exists a finite set
  $\theta$ of tiles over the alphabet $\Sigma \cup \{ \# \}$ such that
  $L = \{ p \in \Sigma^{++} \mid \llbracket \hat{p} \rrbracket \subseteq \theta
  \}$. We will refer to such language as $LOC(\theta)$.

\noindent {\em Locally testable languages} (LT) are analogous to local languages, but
are defined through square tiles with side size possibly bigger than 2. In the rest of the paper
we will call these variant of tiles {\em k-tiles}, to avoid confusion with
standard $2\times2$ tiles. For instance, 3-tiles are square pictures of size (3,3).
\end{definition}

Last, we define {\em tiling systems} (TS). Tiling systems define the
closure w.r.t. alphabetic projection of local languages, and are
presented and studied extensively in \cite{GR-two-dl}.

\begin{definition}\label{DefTS}
  % (Definition 7.2 of \cite{GR-two-dl}) 
  A {\em tiling system} (TS) is a
  4-tuple $\mathcal{T} = (\Sigma, \Gamma, \theta, \pi)$, where
  $\Sigma$ and $\Gamma$ are two finite alphabets, $\theta$ is a
  finite set of tiles over the alphabet $\Gamma \cup \{\#\}$, and $\pi
  : \Gamma \to \Sigma$ is a projection. \\
The language defined by the tiling system $\mathcal{T}$ (in the rest denoted by
$L(\mathcal{T})$) is  the set of pictures $\{ \pi(p) \mid
\hat{p} \in LOC(\theta) \}$.
\end{definition}

\section{Tile grammars}\label{SectionTileGrammars}

We are going to introduce and study a very general grammar type
specified by a set of rewriting rules (or productions). 
A typical rule has a left and a
right part, both pictures of unspecified but equal (isometric)
size. The left part is an $A$-homogeneous picture, where $A$ is a
nonterminal symbol. The right part is a picture of a local language
over nonterminal symbols.  Thus a rule is a scheme defining a possibly
unbounded number of isometric pairs: left picture, right picture. In
addition there are simpler rules whose right part is a single terminal.
\par
The derivation process of a picture starts from a
$S$(axiom)-homogeneous picture.  At each step, an $A$-homogeneous
subpicture is replaced with an isometric picture of the local
language, defined by the right part of a rule $A \to \ldots$. The
process terminates when all nonterminals have been eliminated from the
current picture.  

For simplicity, this presentation focuses on nonterminal rules, thus
excluding for instance that both terminal and nonterminal symbols are
in the same right part.  This normalization has a cost in terms of
grammar dimension and readability, but does not lose
generality. Indeed, more general kinds of rules (e.g. like those used
in \cite{TRG1}), can be easily simplified by introducing some
auxiliary nonterminals and rules. We will present and use analogous
transformations when comparing with other grammar devices in Section
\ref{SectionComparisons}, where we will talk about {\em nonterminal
  normal forms}.

\begin{definition} \label{TG}
  A {\em tile grammar (TG)}
  is a tuple $(\Sigma, N, S, R)$, where $\Sigma$ is the {\em terminal}
  alphabet, $N$ is a set of {\em nonterminal} symbols, $S \in N$ is
  the {\em starting symbol}, $R$ is a set of {\em rules}.
\\
Let $A \in N$. There are two kinds of rules:
\begin{eqnarray}
\text{Fixed size: }&  A \rightarrow t, &\text{ where } t \in \Sigma; \\
\text{Variable size: }&  A \rightarrow \omega,\  & \ \omega \text{ is a
  set of non-concave tiles over } N \cup \{ \# \}.
\end{eqnarray}

\end{definition}

\noindent {\em Concave tiles} are like
$
\begin{array}{|cc|}
\hline
B & B \\
C & B \\
\hline
\end{array}
$ 
or a rotation thereof, where $B \ne \#$ (so we use concave tiles only for
borders). It is easy to see that  
all pictures in $LOC(\omega)$, where $\omega$ is a set of non-concave tiles, 
admit a strong homogeneous partition.

Picture derivation is next defined as a relation between partitioned
pictures.

\begin{definition} \label{derivation} Consider a tile grammar $G = (\Sigma,
  N, S, R)$, let $p,p' \in (\Sigma \cup N)^{(h,k)}$ be pictures of
  identical size. Let $\pi = \{d_1,\ldots,d_n\}$ be a homogeneous partition of
  $\mathrm{dom}(p)$.
  We say that $(p',\pi')$ {\em derives in one step} from $(p,\pi)$,
  written
\[
(p,\pi)\Rightarrow_G (p',\pi')
\]
iff, for some $A \in N$,
there exist in $\pi$ an $A$-homogeneous subdomain $d_i = (x, y; x',
y')$, called \emph{application area}, and a rule $A \to \alpha \in R$
such that
 $p'$ is obtained substituting $ \mathrm{spic}(p,d_i) $ in $p$
  with:
\begin{itemize}
\item $\alpha \in \Sigma$, if $A \to \alpha$ is of type (1);\footnote{In this case, $x=x'$ and $y=y'$.}
\item $s \in LOC(\alpha)$, if $A \to \alpha$ is of type (2).
% and $s$ admits a strong homogeneous partition $\Pi(s)$.
\end{itemize}
Moreover, $\pi' = (\pi \setminus \{d_i\}) \cup
(\Pi(s) \oplus (x-1,y-1))$.
%\item 
%is a homogeneous partition of $\mathrm{dom}(p)$ into the
%  subdomains
%  \[
%  (\pi \setminus \{d_i\}) \cup  \mathrm{transl}_{(x-1,y-1)}(\Pi(s))
%  \]
%\end{itemize}
%\noindent where $\mathrm{transl}_{(x-1,y-1)}(\Pi(s))$ is the
%translation by displacement $(x-1,y-1)$ (intuitively, the position of
%$d_i$ in $p$) of the subdomains of $\Pi(s)$, the homogeneous
%partition of $s$.

\noindent We say that $(p',\pi')$ {\em derives from $(p,\pi)$} in $n$ steps,
written $(p,\pi) \stackrel{n}{\Rightarrow}_G (p',\pi')$, iff $p = p'$
and $\pi=\pi'$, when $n = 0$, or there are a picture $p''$ and a
homogeneous partition $\pi''$ such that $(p,\pi)
\stackrel{n-1}{\Longrightarrow}_G (p'',\pi'')$ and $(p'',\pi'')
\Rightarrow_{G} (p',\pi')$. We use the abbreviation $(p,\pi)
\stackrel{*}{\Rightarrow}_{G} (p',\pi')$ for a derivation with a finite
number of steps. 
\end{definition}

Roughly speaking, at each step of the derivation an $A$-homogeneous
subpicture is replaced with an isometric picture of the local
language, defined by the right part of a rule $A \to \alpha$, that
admits a strong homogeneous partition. The process terminates when
all nonterminals have been eliminated from the current
picture.

In the rest of the paper, and when considering
also other grammatical devices, we will drop the $G$ symbol when it is clear from the
context, writing e.g. $(p,\pi) \stackrel{*}{\Rightarrow} (p',\pi')$.

\begin{definition} \label{piclang}
  The {\em picture language} defined by a grammar $G$ (written $L(G)$)
  is the set of $p \in \Sigma^{++} $ such that
\[
    \left(S^{|p|},\left\{\mathrm{dom}(p)\right\}\right) \stackrel{*}{\Rightarrow}_{G} \left(p,\mathrm{unit}(p)\right)
\]
   For short we also write $S
  \stackrel{*}{\Rightarrow}_{G} p$.
\end{definition}

We emphasize that, to generate a picture of a certain dimension, one
must start from a picture of the same dimension.

We also will use the notation $\LL(X)$ to denote the class of
languages generated by some formal device $X$, e.g. $\LL(TG)$ will
denote the class of languages generated by tile grammars.

The following examples will be used later for comparing language families.
\begin{example}\label{twoBytwoDividerEx}
  \emph{One row and one column of $b$'s}.
  \\
  The set of pictures having one row and one column (both
  not at the border) that hold $b$'s, and the remainder of the picture
  filled with $a$'s is defined by the tile grammar $G_1$ in Figure
  \ref{FigGrammEx1}, where the nonterminals are $\{A_1, A_2, A_3, A_4, V_1, V_2,
H_1, H_2, X, A, B\}$.
\begin{figure}[h!]
{\small
\[
G_1 : \ \ \ 
S \to \left\llbracket
\begin{array}{ccccccc}
    \#  & \#  & \#  & \# & \#  & \# & \# \\
    \#  & A_{1} & A_{1} & V_1 & A_{2} & A_{2} & \#  \\
    \#  & A_{1} & A_{1} & V_1 & A_{2} & A_{2} & \#  \\
    \#  & H_1   & H_1  & V_1  & H_2  & H_2   & \#  \\
    \#  & A_{3} & A_{3} & V_2 & A_{4} & A_{4} & \#  \\
    \#  & A_{3} & A_{3} & V_2 & A_{4} & A_{4} & \#  \\
    \#  & \#  & \#  & \#  & \#  & \# & \# \\
  \end{array}
\right\rrbracket
\]
\[
\ \ A_i \to \left\llbracket
\begin{array}{cccc}
    \#  & \#   & \#  & \# \\
    \#  & X    &  X & \# \\
    \#  & A_i  &  A_i  & \# \\
    \#  & A_i  &  A_i  & \# \\
    \#  & \#   & \#  & \# \\
  \end{array}
\right\rrbracket \mid \left\llbracket
\begin{array}{cccc}
    \#  & \#  & \#  & \# \\
    \#  & X & X & \# \\
    \#  & \#  & \#  & \# \\
  \end{array}
\right\rrbracket, \ \text{for} \ 1 \le i \le 4
\]
\[
X \to \left\llbracket
\begin{array}{ccccc}
    \#  & \#  & \#  & \#  & \# \\
    \#  & A & X & X & \# \\
    \#  & \#  & \#  & \#  & \# \\
  \end{array}
\right\rrbracket \mid a; \ \ H_i \to \left\llbracket
\begin{array}{ccccc}
    \#  & \#  & \#  & \#  & \# \\
    \#  & B   & H_i   & H_i   & \# \\
    \#  & \#  & \#  & \#  & \# \\
  \end{array}
\right\rrbracket \mid b, \ \text{for} \ 1 \le i \le 2
\]
\[
A \to a; \ \ B \to b; \ \ V_i \to \left\llbracket
\begin{array}{ccc}
    \#  & \#  & \# \\
    \#  & B & \# \\
    \#  & V_i & \# \\
    \#  & V_i & \# \\
    \#  & \#  & \# \\
  \end{array}
\right\rrbracket \mid b, \ \text{for} \ 1 \le i \le 2.
\]
\medskip
\[
p_1 = 
  \begin{array}{|ccccc|}
\hline
 a & a & b & a & a \\
 b & b & b & b & b \\
 a & a & b & a & a \\
 a & a & b & a & a \\
\hline
  \end{array}
\]
}
\caption{Tile grammar $G_1$ (top) and a picture $p_1$ (bottom) of Example \ref{twoBytwoDividerEx}.}\label{FigGrammEx1}
\end{figure}
We recall that $\llbracket\,\rrbracket$ denotes the set of tiles
contained in the argument picture. This notation is preferable to the
listing of all tiles, shown next:
\[
S \to \left\{
\begin{array}{|cc|}
\hline
    \#  & \# \\
    \#  & A_1 \\
\hline
\end{array}
,
\begin{array}{|cc|}
\hline
    \#  & \# \\
    A_1  & A_1 \\
\hline
\end{array}
, \ldots,
\begin{array}{|cc|}
\hline
    A_1  & V_1 \\
    H_1  & V_1 \\
\hline
\end{array}
,
\begin{array}{|cc|}
\hline
    V_1  & A_2 \\
    V_1  & H_2 \\
\hline
\end{array},
\ldots,
\begin{array}{|cc|}
\hline
    A_4  & A_4 \\
    \#   & \# \\
\hline
\end{array},
\begin{array}{|cc|}
\hline
    A_4  & \# \\
    \#  & \# \\
\hline
\end{array}
\right\}.
\]
An example of derivation is shown in Figure \ref{FigDerivationEx1}, where
partitions are outlined for readability.
\begin{figure}[h!]
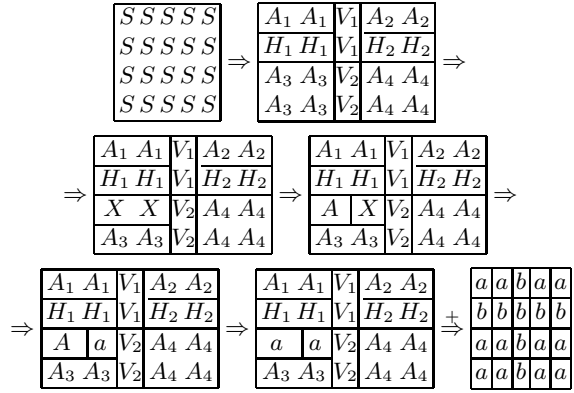

 {\small
\[
\begin{array}{|@{\,}c@{\,}c@{\,}c@{\,}c@{\,}c|}
\hline
 S & S & S & S & S \\
 S & S & S & S & S \\
 S & S & S & S & S \\
 S & S & S & S & S \\
 \hline
  \end{array}
\Rightarrow
\begin{array}{|@{\,}c@{\,}c@{\,}c@{\,}c@{\,}c|}
\hline
   A_{1} & A_{1} &\vline V_1 \vline& A_{2} & A_{2} \\
 \cline{1-2} \cline{4-5}
   H_1   & H_1  &\vline V_1  \vline& H_2  & H_2 \\
 \hline
   A_{3} & A_{3} &\vline V_2 \vline& A_{4} & A_{4} \\
   A_{3} & A_{3} &\vline V_2 \vline& A_{4} & A_{4} \\
\hline
  \end{array}
\Rightarrow
\]
\[
\Rightarrow
\begin{array}{|@{\,}c@{\,}c@{\,}c@{\,}c@{\,}c|}
\hline
   A_{1} & A_{1} &\vline V_1 \vline& A_{2} & A_{2} \\
 \cline{1-2} \cline{4-5}
   H_1   & H_1  &\vline V_1  \vline& H_2     & H_2 \\
\hline
  X     & X     &\vline  V_2 \vline& A_{4} & A_{4} \\
\cline{1-2}
   A_{3} & A_{3} &\vline  V_2 \vline& A_{4} & A_{4} \\
\hline
  \end{array}
\Rightarrow
\begin{array}{|@{\,}c@{\,}c@{\,}c@{\,}c@{\,}c|}
\hline
   A_{1} & A_{1} &\vline V_1 \vline& A_{2} & A_{2} \\
 \cline{1-2} \cline{4-5}
   H_1   & H_1  &\vline V_1  \vline& H_2     & H_2 \\
\hline
  A     &\vline \  X     &\vline  V_2 \vline& A_{4} & A_{4} \\
\cline{1-2}
   A_{3} & A_{3} &\vline  V_2 \vline& A_{4} & A_{4} \\
\hline
  \end{array}
\Rightarrow
\]
\[
\Rightarrow
\begin{array}{|@{\,}c@{\,}c@{\,}c@{\,}c@{\,}c|}
\hline
   A_{1} & A_{1} &\vline V_1 \vline& A_{2} & A_{2} \\
 \cline{1-2} \cline{4-5}
   H_1   & H_1  &\vline V_1  \vline& H_2     & H_2 \\
\hline
  A     &\vline\  a     &\vline  V_2 \vline& A_{4} & A_{4} \\
\cline{1-2}
   A_{3} & A_{3} &\vline  V_2 \vline& A_{4} & A_{4} \\
\hline
  \end{array}
\Rightarrow
\begin{array}{|@{\,}c@{\,}c@{\,}c@{\,}c@{\,}c|}
\hline
   A_{1} & A_{1} &\vline V_1 \vline& A_{2} & A_{2} \\
 \cline{1-2} \cline{4-5}
   H_1   & H_1  &\vline V_1  \vline& H_2     & H_2 \\
\hline
  a     &\vline \ a     &\vline  V_2 \vline& A_{4} & A_{4} \\
\cline{1-2}
   A_{3} & A_{3} &\vline  V_2 \vline& A_{4} & A_{4} \\
\hline
  \end{array}
\overset{+}\Rightarrow
  \begin{array}{|c|c|c|c|c|}
\hline
 a & a & b & a & a \\
\hline
 b & b & b & b & b \\
\hline
 a & a & b & a & a \\
\hline
 a & a & b & a & a \\
\hline
  \end{array}
\]}
\caption{Derivation using grammar $G_1$ of Example
  \ref{twoBytwoDividerEx}, Figure \ref{FigGrammEx1}, with
 outlined partitions.}
\label{FigDerivationEx1}
\end{figure}
\end{example}

\begin{example}\label{PalindromicRowsEx}
  \emph{Pictures with palindromic rows}. Each row is an even
  palindrome over $\{a,b\}$. The grammar $G_2$ is shown in Figure
  \ref{FigGrammEx2}.

\begin{figure}[h!]
{\small
\[
G_2 : \ \ \ 
 S_{P}  \to  \left\llbracket
\begin{array}{ccccccc}
    \#  & \#  & \#   & \# \\
    \#  & R & R &   \#  \\
    \#  &  S_{P} & S_{P} &   \#  \\
    \#  & S_{P} & S_{P} &   \#  \\
    \#  & \#  & \#  & \#\\
  \end{array}
\right\rrbracket \mid \left\llbracket
\begin{array}{cccc}
    \#  & \#  & \#   & \# \\
    \#  & R & R &   \#  \\
    \#  & \#  & \#  & \#\\
  \end{array}
\right\rrbracket
\]
\[
  R \to  \left\llbracket
  \begin{array}{cccccc}
    \#  & \#  & \# & \#  & \#  & \#\\
    \#  & A & R & R & A' &  \#  \\
    \#  & \#  & \# & \#  & \# & \#\\
  \end{array}
\right\rrbracket \mid \left\llbracket
  \begin{array}{cccccc}
    \#  & \#  & \# & \#  & \#  & \#\\
    \#  & B & R & R & B' &  \#  \\
    \#  & \#  & \# & \#  & \# & \#\\
  \end{array}
\right\rrbracket
\]
\[
  R \to
\left\llbracket
  \begin{array}{cccc}
    \#  & \#  & \#  & \#\\
    \#  & A & A' &  \#  \\
    \#  & \#  & \#  & \# \\
  \end{array}
\right\rrbracket \mid \left\llbracket
  \begin{array}{cccc}
    \#  & \#  & \#  & \#\\
    \#  & B & B' &  \#  \\
    \#  & \#  & \#  & \# \\
  \end{array}
\right\rrbracket
\]
\[
A  \to a ; \ \ B  \to b ; \ \ A' \to a ; \ \ B' \to b.
\]
\medskip
\[
p_2 = 
  \begin{array}{|cccc|}
\hline
 a & b & b & a \\
 b & a & a & b \\
 a & a & a & a \\
\hline
  \end{array}
\]
}
\caption{Tile grammar $G_2$ (top) and a picture $p_2$ (bottom) of Example \ref{PalindromicRowsEx}.}\label{FigGrammEx2}
\end{figure}
\end{example}

\subsection{Properties of tile grammars}

First, we state a language family inclusion between tiling
systems (Definition \ref{DefTS}) and tile grammars, proved in
\cite{TRG1}. We will illustrate it with an example, both to give the reader
an intuitive idea of the result, and to later re-use the example.

\begin{proposition}\label{PropTSincludedTG}
$\LL(TS) \subset \LL(TG)$.
\end{proposition}

Consider a TS $T = (\Sigma, \Gamma, \theta, \pi)$, where $\Sigma$ is
the terminal alphabet, $\theta$ is a tile-set, $\Gamma$ is the
tile-set alphabet, and $\pi : \Gamma \to \Sigma$ is an alphabetic
projection. It is quite easy to define a TG $T'$ such that $L(T') =
L(T)$. Informally, the idea is to take the tile-set $\theta$ and add
two markers, e.g. $\{b,w\}$ in a ``chessboard-like'' fashion to
build up a tile-set suitable for the right part of the variable size
starting rule; other straightforward fixed size rules are used to
encode the projection $\pi$.  

%The inclusion is proper: in \cite{TRG1}
%a variant of Example \ref{PalindromicRowsEx} is proved not to be in
%$\LL(TS)$. In another way, when considering 1D picture languages
%(e.g. pictures having only one row), $\LL(TS)$ coincides with regular
%languages, while $\LL(TG)$ defines context-free languages.

We note how  both $\LL(TS)$ and $\LL(TG)$ are closed under intersection with the
class of all height-1 pictures: the classes resulting in that intersection are
the well-known classes recognizable and context-free, respectively, string
languages. The inclusion is hence proper: any context-free, non-recognizable
string language is also (when considered as a picture language)
in $\LL(TG)$, but not in $\LL(TS)$.

The next example illustrates the reduction from a TS to a % non-regional
TG.
\begin{example}\label{ex:ts}Square pictures of $a$'s.
  \\
  The TS $T_3$ is based on a local language over $\{0,1\}$ such that
  all pixels of the main diagonal are 1 and the remaining ones are
  0, and on the projection $\pi(0) = \pi(1) = a$.  $T_3$ and the
  equivalent TG $G_3$ are shown in Figure \ref{FigEx3}.

  The ``chessboard-like'' construction is used to ensure that the only
  strong homogeneous partition obtained in applying a rule is the one
  in which partitions correspond to single pixels. This allows the
  application of terminal rules encoding projection $\pi$.
  Note that in the first rule of grammar $G_3$ we used tiles arising
  from the two possible chessboard structures, i.e. the one with a
  ``black'' in top-left position, and the one with a ``white'' in the
  same place. Indeed, to fill areas above and below the diagonal
  with 0's we need both tiles
  \[
  \begin{array}{|cc|}
\hline
    0_b & 0_w  \\
    0_w & 0_b \\
\hline
  \end{array}
 \ \text{ and } \ 
  \begin{array}{|cc|}
\hline
    0_w & 0_b  \\
    0_b & 0_w \\
\hline
  \end{array}.
\]

%non possiamo citare le regionali: non sono state ancora introdotte! 
%Note that $G_3$ is not regional.
\begin{figure}[h!]
\begin{center}
%Tiling system $T_3$:
\[
T_3: \ \ \ \theta = \left\llbracket
  \begin{array}{cccccc}
    \#  & \#  & \#  & \# & \# & \# \\
    \#  &  1  &  0  &  0 & 0 & \# \\
    \#  &  0  &  1  &  0 & 0 & \# \\
    \#  &  0  &  0  &  1 & 0 & \# \\
    \#  &  0  &  0  &  0 & 1 & \# \\
    \#  & \#  & \#  & \# & \# & \# \\
  \end{array}
\right\rrbracket, \ \ \pi(0) = a, \ \ \pi(1) = a.
\]
%Equivalent tile grammar $G_3$:
\[
G_3: \ \ \ S \to \left\llbracket
  \begin{array}{cccccc}
    \#  & \#  & \#  & \# & \# & \# \\
    \#  &  1_b  &  0_w  &  0_b & 0_w & \# \\
    \#  &  0_w  &  1_b  &  0_w & 0_b & \# \\
    \#  &  0_b  &  0_w  &  1_b & 0_w & \# \\
    \#  &  0_w  &  0_b  &  0_w & 1_b & \# \\
    \#  & \#  & \#  & \# & \# & \# \\
  \end{array}
\right\rrbracket \cup \left\llbracket
  \begin{array}{cccccc}
    \#  & \#  & \#  & \# & \# & \# \\
    \#  &  1_w  &  0_b  &  0_w & 0_b & \# \\
    \#  &  0_b  &  1_w  &  0_b & 0_w & \# \\
    \#  &  0_w  &  0_b  &  1_w & 0_b & \# \\
    \#  &  0_b  &  0_w  &  0_b & 1_w & \# \\
    \#  & \#  & \#  & \# & \# & \# \\
  \end{array}
\right\rrbracket
\]
\[
1_w \to a, \ \ 1_b \to a, \ \ 0_w \to a, \ \ 0_b \to a.
\]
\end{center}
\caption{For Example \ref{ex:ts} the TS defining 
$\{a^{(n,n)} \mid n > 1\}$ (top), and the equivalent TG grammar (bottom).
%The TS $T_3$ (top) and the equivalent TG $G_3$ (bottom) for
%  Example \ref{ex:ts}.
}\label{FigEx3}
\end{figure}

\end{example}

The following complexity property will be used to separate the TG
language family from several subfamilies to be introduced.

In this paper as ``parsing problem'' we consider the problem of
deciding if a given input picture is in $L(G)$, for a fixed grammar
$G$ (i.e. the also called {\em non-uniform membership problem}). 
The complexity of parsing algorithms is thus expressed in term
of the size of the input string, in this case the picture size.
 
\begin{proposition}\label{parse-tg-is-np}
    The parsing problem for $\LL(TG)$ is NP-complete.
\end{proposition}

\begin{pf}
  From Proposition \ref{PropTSincludedTG} and the fact that the
  parsing problem for $\LL(TS)$ is NP-complete (see \cite{lindgren}
  where tiling systems are called {\em homomorphisms of local lattice
  languages}, or \cite{lewis}) it follows that parsing $\LL(TG)$ is
  NP-hard.
  \\
  For NP-completeness, we show that parsing $\LL(TG)$ is in NP.
  First, we assume without loss of generality that a TG $G$ does not
  contain any chain rule, i.e. a rule of the form
\[
A \to \left\llbracket
\begin{array}{cccc}
    \#  & \#   & \#  & \# \\
    \#  & B  &  B  & \# \\
    \#  & B  &  B  & \# \\
    \#  & \#   & \#  & \# \\
  \end{array}
\right\rrbracket, \ \ \  B \in N
\]
that corresponds to a renaming rule of a string grammar.

If this is not the case, it is possible to discard chain rules by
directly using the well-known (e.g. \cite{Harrison78}) approach for
context-free string grammars.
\\
We suppose to have a candidate derivation
    \[\left(S^{(h,k)},\mathrm{dom}(p)\right) \Rightarrow_G (p_1, \pi_1)
    \Rightarrow_G (p_2, \pi_2) \Rightarrow_G \cdots \Rightarrow_G
    (p_{n-1}, \pi_{n-1}) \Rightarrow_G \left(p,
      \mathrm{unit}(p)\right) \] and we are going to prove that
    checking its correctness takes polynomial time in $h,k$ (size of
    the picture), by considering the dominant parameters of time complexity.
    \\
    First, the length $n$ of this derivation, since there are no chain
    rules, is at most $h \cdot k$. In fact, we start from a partition
    with only one element coinciding with $\mathrm{dom}(p)$, and at
    each step at least one element is added, arriving at step $n$,
    where the number of elements is $h \cdot k$, each corresponding to
    a pixel.
    \\
    For each step, we must find the application area in $(p_i,\pi_i)$,
    and the corresponding rewritten nonterminal $A$, by comparing
    $(p_i,\pi_i)$ with $(p_{i+1},\pi_{i+1})$. The number of comparisons to be
    performed is at most $h \cdot k$.
    \\
    Then, we have to find a rule $A \to \omega$ in
    $R$ which is compatible with the rewritten subpicture of $p_{i+1}$
    corresponding to the application area. So, at most we must check
    every rule in $R$, and every tile of its right part, on a
    subpicture, given by the application area, which is at most $h
    \cdot k$. Hence, we have to consider for this step a number of checks that
    is at most
    \[
     h \cdot k \cdot |R| \cdot \underset{A \to \omega \in
    R}{\mathrm{max}}|\omega|
    \]
    Each of these considered steps can be done in polynomial time in every
    reasonable machine model, hence the resulting time complexity is still
    polynomial.  
   % We already know that the derivation length is $O(h \cdot k)$,
   % therefore the time complexity of the derivation check is
   % \[
   % O\left( h^2 \cdot k^2 \cdot |R| \cdot \underset{A \to \omega \in
   % R}{\mathrm{max}}|\omega|    \right)
   % \]
    \qed
\end{pf}

From \cite{TRG1} it is known that the family of TG languages is closed
w.r.t. union, column/row concatenation, column/row closure operations,
rotation, and alphabetic mapping.

We mention that all the families presented in this work, that exactly
define the context-free string languages if restricted to one
dimension (i.e. all but tiling systems and grid grammars, presented in
Section \ref{sec:grid}), are not closed w.r.t. intersection and
complement.  This is proved as for string context-free languages: it
is straightforward to see that they are all closed w.r.t. union. But
it is well known that the language $\{a^n b^n c^n \mid n>0\}$ is not
context-free, and can be expressed as intersection of two context free
languages, e.g. $\{a^n b^m c^n \mid m,n>0\}$ and $\{a^n b^n c^m \mid
m,n>0\}$. Hence, they are not closed w.r.t. intersection, but this
also means that they are not closed w.r.t. complement.

%%% Local Variables: 
%%% mode: latex
%%% TeX-master: "paper"
%%% End: 

\section{Regional tile grammars}\label{SectionRegionalGrammars}

We now introduce the central concept of {\em regional language}, and
a corresponding specialization of tile grammars.
 The adjective ``regional'' is a
metaphor of geographical political maps, where different regions
are filled with different colors; of course, regions are rectangles.

Regional tile grammars are central to this work, because they are the
most general among the polynomial-time parsable grammar models
considered in this paper. We will see that it is easy to define
the other kinds of 2D grammars by restricting the tiles
used in regional tile grammars.

\begin{definition} \label{regionalpartition} A homogeneous partition
  is {\em regional} (HR) iff distinct (not necessarily adjacent)
  subdomains have distinct labels. A picture $p$ is {\em
    regional} if it admits a HR partition.
  A language is {\em regional} if all its pictures are so.
\end{definition}

For example, consider Figure \ref{example-partitions}: the partitions
in subdomains of the picture on the left is homogeneous and strong,
but not regional, since four different subdomains bear the same symbol
$A$. On right, a variant of the same picture with regional partitions
outlined is depicted.

\begin{figure}[h!]
\[
\begin{array}{|@{\ }c@{\ }c@{\ }c@{\ }c@{\ }c|}
\hline
   A & A &\vline B \vline& A & A \\
   A & A &\vline B \vline& A & A \\
 \cline{1-2} \cline{4-5}
   D   & D  &\vline B  \vline& D  & D \\
 \hline
   A & A &\vline C \vline& A & A \\
   A & A &\vline C \vline& A & A \\
\hline
  \end{array}
\ \ \ \ \ 
\begin{array}{|@{\ }c@{\ }c@{\ }c@{\ }c@{\ }c|}
\hline
   A_{1} & A_{1} &\vline B \vline& A_{2} & A_{2} \\
   A_{1} & A_{1} &\vline B \vline& A_{2} & A_{2} \\
 \cline{1-2} \cline{4-5}
   D_1    &  D_1  &\vline B  \vline& D_2  & D_2 \\
 \hline
   A_{3} & A_{3} &\vline C \vline& A_{4} & A_{4} \\
   A_{3} & A_{3} &\vline C \vline& A_{4} & A_{4} \\
\hline
  \end{array}
\]
\caption{Pictures with outlined partitions in subdomains: strong homogeneous
  partition (left), and regional (right).}\label{example-partitions}
\end{figure}

Another (negative) example is in Figure \ref{FigEx3}:
``chessboard-like'' pictures admit unique homogeneous partitions,
i.e. those in which every subdomain corresponds to a single
pixel. Note that in general these partitions are strong (adjacent
subdomains have different symbols, like in a chessboard), but are not
regional (e.g. in the variable size rule of grammar $G_3$ there are
multiple $0_b$ symbols).

\begin{definition} \label{RTG}
  A {\em regional tile grammar (RTG)} is a tile grammar (see Definition
  \ref{TG}), in which every variable size rule $A \to \omega$
  is such that $LOC(\omega)$ is a {\em regional language}.
%   is a tuple $(\Sigma, N, S, R)$, where $\Sigma$ is the {\em terminal}
%   alphabet, $N$ is a set of {\em nonterminal} symbols, $S \in N$ is
%   the {\em starting symbol}, $R$ is a set of {\em rules}.
% \\
% Let $A \in N$. There are two kinds of rules:
% \begin{eqnarray*}
% \text{Fixed size: }&  A \rightarrow t, &\text{ where } t \in \Sigma; \\
% \text{Variable size: }&  A \rightarrow \omega,\  & \ \omega \text{ is a
%   set of tiles over } N \cup \{ \# \},\\
% & & LOC(\omega) \text{ is a regional language.}
% \end{eqnarray*}
\end{definition}

We note that the tile grammars presented in Examples
\ref{twoBytwoDividerEx} and \ref{PalindromicRowsEx} are regional,
while the one of Example \ref{ex:ts} ($G_3$) is not. Another
RTG is presented in the following example.

% The example can be generalized to pictures composed of $m\times m$
% rectangular checkerboard-like fields, delimited by intersecting two
% systems of $m-1$ vertical and horizontal straight lines.
%%%%%%%%%%%%%%%%%%%%%%

%%%%%%%%%%%%%%%%%%%%%%%%%%%%%%%%%%%%
\begin{example}\label{MisalignedPalindromicRowsEx}
    \emph{Misaligned  palindromes}. 

    A picture is a ``ribbon'' of two rows, divided into four fields:
    at the top-left and at the bottom right of the picture are
    palindromes as in Example \ref{PalindromicRowsEx} (where rules for $S_p$ are
    defined). The other two fields are filled with $c$'s and must not
    be adjacent. The corresponding regional tile grammar $G_4$ is shown in
    Figure \ref{FigGrammEx4}.

\begin{figure}[h!]
{\small
\[
G_4 : \ \ \ 
 S \to 
\left\llbracket
\begin{array}{cccccccc}
    \#  & \#  & \#  & \# & \# & \#  & \#  & \# \\
    \#  & P_{1} & P_{1} & P_{1} & P_{1} & C_1 & C_1 &  \#  \\
    \#  & C_2 & C_2 & P_{2} & P_{2} & P_{2} & P_{2} & \#  \\
    \#  & \#  & \#  & \# & \#  & \#  & \#  & \#\\
  \end{array}
\right\rrbracket;
\ \ 
P_{1} \to S_P ; \ \ P_{2} \to S_P
\]
\[
C_i \to
\left\llbracket
\begin{array}{ccccc}
    \#  & \#  & \#  & \#  & \# \\
    \#  & C   & C_i   & C_i   & \# \\
    \#  & \#  & \#  & \#  & \# \\
  \end{array}
\right\rrbracket
\mid
c, \ \text{for} \ 1 \le i \le 2; \  \ \ 
C \to c.
 \]
\medskip
\[
p_4 = 
  \begin{array}{|cccccccccc|}
\hline
 a & a & b & b & a & a & c & c & c & c \\
 c & c & b & a & b & a & a & b & a & b \\
\hline
  \end{array}
\]
}
\caption{Regional tile grammar $G_4$ (top) and a picture $p_4$
  (bottom) of Example
  \ref{MisalignedPalindromicRowsEx}.}\label{FigGrammEx4}
\end{figure}
\end{example}

Next, we study the form of tiles
occurring in a regional local language.

Consider a tile set $\theta$ over the alphabet $\Sigma \cup \{ \#
\}$. For a tile $t$ we define the \emph{horizontal and vertical
  adjacency relations} $\mathcal{H}_t,\mathcal{V}_t \subseteq \left(
  \Sigma \cup \{ \# \} \right)^2$ over its pixels $t(i,j)$ as
\[
\forall i, \ 1 \le i \le 2, \ t(i,1) \ne t(i,2) \Leftrightarrow t(i,1) \ \mathcal{H}_t \ t(i,2); 
\]
\[
\forall j, \ 1 \le j \le 2, \ t(1,j) \ne t(2,j) \Leftrightarrow t(1,j) \ \mathcal{V}_t \ t(2,j).
\]
Then, the {\em adjacency relations} are 
$\mathcal{A}_t=\mathcal{H}_t \cup \mathcal{V}_t$ and
$\mathcal{A}'_t=\mathcal{H}_t^{-1} \cup \mathcal{V}_t$.

\noindent
The relations can be extended to a tile set $\theta$: $x
\mathcal{H}_{\theta} y$ iff $\exists t \in \theta : x \mathcal{H}_t
y$; and similarly for $\mathcal{V}_{\theta}$,
$\mathcal{A}_{\theta}$, and $\mathcal{A}'_{\theta}$.

\begin{proposition}\label{distRecTileSet}
Let $p \in \Sigma^{++}$ and $\theta = \llbracket \hat{p} \rrbracket$; 
picture $\hat{p}$ is regional iff
%\begin{enumerate}
%\item the (finite) language $\theta \cap \Sigma^{(2,2)}$ (which
%  excludes the tiles containing \#) is regional,
%  and 
%\item 
the incidence graphs of both 
$\mathcal{A}_{\theta} \cap \Sigma^2$ and
$\mathcal{A}'_{\theta} \cap \Sigma^2$ are acyclic.
%\end{enumerate}
\end{proposition}

\noindent We will call {\em simple regional} such a tile set.

\begin{pf}
First of all, we note that tiles occurring in regional
pictures have the following form (or a rotation thereof):
\[
\begin{array}{|cc|}
\hline
A & A \\
A & A \\
\hline
  \end{array},
\quad
\begin{array}{|cc|}
\hline
A & A \\
B & B \\
\hline
  \end{array},
\quad
\begin{array}{|cc|}
\hline
A & A \\
B & C \\
\hline
  \end{array},
\quad
\begin{array}{|cc|}
\hline
A & B \\
C & D \\
\hline
  \end{array},
\quad
\begin{array}{|cc|}
\hline
\# & \# \\
A  & \# \\
\hline
  \end{array},
\quad
\begin{array}{|cc|}
\hline
\# & \# \\
A  & A  \\
\hline
  \end{array},
\quad
\begin{array}{|cc|}
\hline
\# & \# \\
A  & B  \\
\hline
  \end{array},
\]
with $A,B,C,D$ all different. The incidence graphs of
the adjacency relations of this tile-set are clearly all acyclic. Moreover, a picture exclusively made of these
kind of tiles admits a unique strong homogeneous partition.
So, if we start from a regional picture $\hat{p}$, we obtain acyclic incidence
graphs for the tile-set made of all its tiles.

Vice versa, if we consider a tile set $\theta$ such that its adjacency relations
are both acyclic, then tiles in $\theta$ must be like those considered in the
previous paragraph. Also, for any picture in $LOC(\theta)$, an acyclic 
$\mathcal{A}_\theta$ means that any path going from the top-left
corner and arriving to the bottom-right corner and performing only down
and right movements cannot traverse two distinct subdomains bearing the same
label. 
For $\mathcal{A'}_{\theta}$ it is analogous, but starting
from the top-right corner, arriving to the bottom-left corner and
performing only left and down movements.
But this means that $LOC(\theta)$ is a regional language.
\qed

% Intuitively, a cyclic incidence graph of $\mathcal{A}_{\theta}$ means
% that the strong homogeneous partition of $p$ contains two distinct
% subdomains with identical labels, and that it is possible to find such
% areas starting from the top-left corner of $p$, arriving to the
% bottom-right corner and following a path consisting of right and down
% movements. 
% More precisely, let the cycle be $B_0=A,B_1,B_2,\ldots, B_a,A=B_{a+1}$. This
% means that, for each step $(B_{j},B_{j+1})$ of the path there is at
% least a tile $t_j$ which is present in $p$ and such that $B_{j}$ and
% $B_{j+1}$ are horizontally or vertically adjacent, hence it is
% possible to reach a $B_{j+1}$-labeled partition from a $B_{j}$-labeled
% partition with a right or down movement, respectively. Being the strong
% homogeneous partition unique, this also means that, following the
% cycle, it is possible to exit a $A$-labeled partition and to reach another 
% one bearing the same label, thus breaking the regionality condition.
% 
% For $\mathcal{A'}_{\theta}$ the construction is analogous, but starting
% from the top-right corner, arriving to the bottom-left corner and
% performing only left and down movements.
% 
% If $p$ is regional, then it is impossible to find such areas, thus both
% graphs are acyclic. Vice versa, if both graphs are acyclic,
% then it is impossible to find a path which traverses identically
% labeled areas of the strong homogeneous partition, hence $p$ is
% regional.  
\end{pf}

\begin{proposition}\label{regTileSet}
  A local language $L$ is regional iff there exist some simple regional tile
  sets $\theta_1,$ $\theta_2,$ $\ldots,$ $\theta_n$, $n \ge 1$,
  such that $L = \bigcup_{1 \le i \le n} LOC(\theta_i)$.
\end{proposition}

\begin{pf} 
If $\theta$ is not simple regional, then it is possible to find a
cycle in one of the incidence graph, let it be $A, B_1, B_2 \ldots
B_a, A$. We consider now all the tiles determining each edge of the
cycle (e.g. for the first step of the cycle, all tiles containing an
$A$ and a $B_1$ that are vertically or horizontally adjacent). Call
such tiles $t_1, t_2, \ldots, t_{b}$, with $b \ge a+1$.
Clearly, the tile sets $\theta_i = \theta \setminus \{ t_i \}$, $1
\le i \le b$, are such that $\bigcup_{1 \le i \le b} LOC(\theta_i)
\subseteq LOC(\theta)$.  Let us suppose that there exists a picture
$p_{b}$ in $LOC(\theta)$ containing all the tiles $t_i$, $1 \le i
\le b$ . But this means that $\llbracket \hat{p_{b}} \rrbracket$ is
not simple regional, because by construction the tiles $t_i$
determine a cycle on one of the incidence graphs of the adjacency
relations, so $p_{b}$ is not regional. Hence,
$LOC(\theta) = \bigcup_{1 \le i \le b} LOC(\theta_i)$.

Now let us consider the tile sets $\theta_i$; if they are all simple
regional, we are done. If not, we repeat the same construction,
until we are able to find the desired $\theta'_1,$ $\theta'_2,$
$\ldots,$ $\theta'_n$. The procedure always terminates, since
$\theta$ is finite. \qed
\end{pf}

Thanks to this result and without loss of generality\footnote{$X \to
  \theta$ generates the same language as the rules $X \to
  \theta_1 \mid \theta_2 \mid \ldots \mid \theta_n$.}, in the rest of
the paper we will always consider regional tile grammar were the right
parts of type (2) rules are simple regional. In practice, right parts will be
written as $\llbracket q \rrbracket$, where $q$ is a bordered regional
picture.

\subsection{Parsing for regional tile grammars}\label{SectionParsingRTG}

To present our version of the Cocke-Kasami-Younger (CKY) algorithm \cite{CKY},
we have to generalize from substrings to subpictures.  Like the CKY algorithm
for strings, our algorithm works bottom-up, by considering all subpictures of
the input picture, starting from single pixels (i.e. $1 \times 1$ subpictures),
and then increasing their size.  As a substring is identified by the positions
of its first and last characters, a subpicture is conveniently identified by
its subdomain. For simplicity and without loss of generality, we assume that
the regional tile grammar considered does not contain variable size chain rules.

The algorithm's main data structure is the {\em recognition matrix}, a
four-dimensional matrix, holding lists of nonterminals, that the
algorithm fills during its run. 
A nonterminal $A$ is put into the matrix entry corresponding to 
subdomain $d$, if the same nonterminal can derive the subpicture
$\mathrm{spic}(p, d)$.

To decide if a rule  can be used to derive the
subpicture corresponding to subdomain $d$, the right part of the rule is
examined, together with all the subdomains contained in
$d$. Type (1) rules are easily managed, because they can only generate
single terminal pixels, therefore they are considered only at the beginning
with unitary subdomains. For example, let us consider grammar
$G_1$ of Example \ref{twoBytwoDividerEx} (Figure \ref{FigGrammEx1}),
and its derivation shown in Figure \ref{FigDerivationEx1}. The pixel
at position $(3,2)$ is an $a$, and the only possible generating
terminal rules are $X \to a$ and $A \to a$. So we enter both $X$ and $A$ into
the recognition matrix at $(3,2;3,2)$.

For type (2) rules we need to check all the pictures in $LOC(\omega)$,
isometric to the considered subpicture.  Thanks to the regional
constraint, every nonterminal used in the right part of the rule
corresponds to a unique homogeneous rectangular area, if the rule is
applicable. So we examine all the sets of nonterminals stored in the
recognition matrix for all the subdomains contained in $d$: if we are
able to find a set of subdomains which comply with the adjacency
relations of the right part of the rule, then the rule is applicable.
For example, let us consider the subdomain $(3,1;3,2)$ for the
derivation of Figure \ref{FigGrammEx1}. Subdomains $(3,1;3,1)$ and
$(3,2;3,2)$ have already been considered, being ``smaller'', and the
set $\{A,X\}$ has been entered at positions $(3,1;3,1)$ and
$(3,2;3,2)$.  This means that, if we consider $X$ at $(3,1;3,1)$, and
$A$ at $(3,2;3,2)$, then all the adjacency relations of the type (2)
rule for $X$ in Figure \ref{FigGrammEx1} are satisfied (namely, $\#
\HH A$, $A \HH X$, $X \HH \#$, $\# \VV A$, $A \VV \#$, $\# \VV X$, $X
\VV \#$).  So the algorithm places $X$ into $(3,1;3,2)$, since
subpicture $(3,1;3,2)$ can be parsed to $X$.

\medskip
\noindent 
{\bf Remark}: {\it 
in the pseudo-code, loops over sets that are Cartesian product are
to be performed in lexicographic order. For example, when stated e.g. 

{\bf for each} $(i, j) \in \{ 1, \ldots, 10\} \times \{ 3, 5, \ldots, 11\}$: 
\ldots

\noindent 
the control variables of the loop (i.e. $i$ and $j$ in this case) will
respectively assume the following sequence of values
in turn:
$(1,3), (1, 5), \ldots, (1, 11), (2, 3), (2, 5), \ldots, (10,11)$. 
}

\smallskip

We now present the details of the algorithm.
Let $p$ be a picture of size $(m,n)$, to be parsed with a
regional tile grammar $G = (\Sigma,N,S,R)$.

\begin{definition}
  A {\em recognition matrix} $\mathfrak{M}$ is a 4-dimensional $m
  \times n \times m \times n$ matrix over the powerset of $N$.
%, whose generic element
%  $\mathfrak{M}(i, j; h, k)$ is a set of nonterminals. 
\end{definition}

Being a generalization of the CKY algorithm for string, the meaning of $A \in \mathfrak{M}(i, j; h, k)$ is that $A$ can derive
the subpicture $\mathrm{spic}(p,(i, j; h, k))$.  In fact, only
cells $(i, j; h, k)$, with $h \ge i, k \ge j$, are used: these cells
are the four-dimensional counterpart of the upper triangular matrix
used in classical CKY algorithm.

We introduce another data structure, the {\em subdomains
  vector}, to be used for recognizing the applicability of
type (2) rules. 

\begin{definition}
  Consider a recognition matrix $\mathfrak{M}$, and a subdomain $d =
  (i,j; k,l)$. Let the nonterminal set $N$ be arbitrarily ordered as
  $A_1, A_2,$ $\ldots,$ $A_{|N|}$. The {\em subdomains vector}
  $\mathfrak{D}(d, \mathfrak{M})$ is the Cartesian product $D_1 \times
  D_2 \times \ldots \times D_{|N|}$, where every $D_t$ is the set of
  subdomains $d'$ such that $A_t \in \mathfrak{M}(d')$ and $d'$ is a
  subdomain contained in $d$; if $D_t$ is empty, then its conventional
  value is set to $(0,0;0,0)$.

  For any nonterminal $A$, the notation $\mathfrak{D}(d,
  \mathfrak{M})|_A$ denotes the component of the vector
  corresponding to $A$.
\end{definition}

To simplify the notation, we shall write $\mathfrak{D}(d)$ instead of
$\mathfrak{D}(d, \mathfrak{M})$ at no risk of confusion, because the
algorithm refers to a unique recognition matrix $\mathfrak{M}$. 

The main role of this ancillary data structure is to assign all the
subdomains contained in a given subdomain $d$, to nonterminals, if
possible, by considering the already filled portion of
$\mathfrak{M}$. Using $\mathfrak{D}$, we are able to check if the
adjacency relations of rules are satisfied. For example, if a rule $A
\to \alpha$ demands $A_2 \mathcal{H}_\alpha A_8$, then we only have to
check if one of the elements of $\mathfrak{D}(d)$ has components 2 and
8 that are horizontally adjacent, with the domain corresponding to
nonterminal $A_2$ to the left.  Figure \ref{computev} shows the
procedure used to compute vector $\mathfrak{D}$.

It is important to remark that $\mathfrak{D}$ is central for keeping the time of
the parsing algorithm polynomial w.r.t. the input size. Indeed, in a regional
tile grammar the number of possible homogeneous subdomains to be considered for
a candidate application area is at most $|N|$, because the number of used
``colors'' in the right part of a rule is at most the number of nonterminals of
the grammar, and when we are considering each element of $\mathfrak{D}$, we know
that it has size less than $(m^2 n^2)^{|N|}$.  In principle, it would be
possible to adapt this algorithm also to an unrestricted tile grammar, but in
this case the number of elements to be considered could be exponential, as the
number of different homogeneous subdomains could be at most as big as the number
of pixels of the application area (see e.g. grammar $G_3$ in
Figure~\ref{FigEx3}).

\begin{figure}
\noindent {\bf Procedure} {\it Compute$\mathfrak{D}$}($\mathfrak{M}$, (i, j; k, l)): \\
\noindent
Every set in $\mathfrak{D}$ is empty;\\
{\bf for each} $(i', j') \in \{ i, \ldots,  k\} \times \{ j, \ldots, l\}$:
\begin{itemize}
\item[] {\bf for each} $(k',l') \in \{i', \ldots, k\} \times \{j', \ldots, l\}$:
\begin{itemize}
\item[] {\bf for each} $A \in \mathfrak{M}(i',j'; k',l')$:
\begin{itemize}
\item[] put $(i',j'; k',l')$ into the set $\mathfrak{D}|_A$;
\end{itemize}
\end{itemize}
\end{itemize}

{\bf for each} $A \in N$:
\begin{itemize}
\item[] {\bf if} $\mathfrak{D}|_A = \emptyset$ {\bf then} 
     put $(0,0;0,0)$ into the set $\mathfrak{D}|_A$;
\end{itemize}

{\bf return} $\mathfrak{D}$.
\caption{Compute$\mathfrak{D}$}\label{computev}
\end{figure}

The actual procedure for checking if a rule of the grammar can be
applied to a given rectangle $(i,j; k,l)$ is presented in Figure
\ref{check-rule}. Based on the vector $\mathfrak{D}$, computed for the
relevant subdomain $(i,j; k,l)$, the procedure checks, for a right
part $\omega$ of a variable-size rule, if all adjacency constraints
are satisfied.

\begin{figure}
\noindent {\bf Procedure} {\it CheckRule} $\left(\mathfrak{D}, \omega, (i,j; k,l) \right)$ : \\
{\bf for each} $(d_1, d_2, \ldots, d_{|N|}) \in \mathfrak{D}$;
\begin{itemize}
\item[] $f := True$; \item[] {\bf for each} $(N_a,N_b) \in \mathcal{H}_\omega$:
\begin{itemize}
\item[] {\bf if} $d_a = (i_a,j_a; k_a,l_a)$ and
$d_b = (i_b,j_b; k_b,l_b)$ are not such that \\
$j_b = l_a+1$, and $k_b \ge i_a, k_a \ge i_b$,\\
{\bf then} $f := False$;
\end{itemize}
\item[] {\bf for each} $(N_a,N_b) \in \mathcal{V}_\omega$:
\begin{itemize}
\item[] {\bf if} $d_a = (i_a,j_a; k_a,l_a)$ and
$d_b = (i_b,j_b; k_b,l_b)$ are not such that \\
$i_b = k_a+1$, and $l_b \ge j_a, l_a \ge j_b$,\\
{\bf then} $f := False$;
\end{itemize}
\item[] {\bf for each} $(\#,N_a) \in \mathcal{H}_\omega$:
\begin{itemize}
\item[] {\bf if} $d_a = (i_a,j_a; k_a,l_a)$ and $j_a \ne j$ {\bf then} $f := False$;
\end{itemize}
\item[] {\bf for each} $(N_a,\#) \in \mathcal{H}_\omega$:
\begin{itemize}
\item[] {\bf if} $d_a = (i_a,j_a; k_a,l_a)$ and $l_a \ne l$ {\bf then} $f := False$;
\end{itemize}
\item[] {\bf for each} $(\#,N_a) \in \mathcal{V}_\omega$:
\begin{itemize}
\item[] {\bf if} $d_a = (i_a,j_a; k_a,l_a)$ and $i_a \ne i$ {\bf then} $f := False$;
\end{itemize}
\item[] {\bf for each} $(N_a,\#) \in \mathcal{V}_\omega$:
\begin{itemize}
\item[] {\bf if} $d_a = (i_a,j_a; k_a,l_a)$ and $k_a \ne k$ {\bf then} $f := False$;
\end{itemize}
\item[] {\bf if} $f$ {\bf then return} $True$;
\end{itemize}
{\bf return} $False$.\\
\caption{CheckRule}\label{check-rule}
\end{figure}

The {\em Main} procedure, presented in Figure \ref{main}, is
structured as a straightforward generalization to two dimensions of
the CKY parsing algorithm. The input picture $p$ is in $L(G)$ iff $S
\in \mathfrak{M}(1,1;m,n)$.

\begin{figure}
\noindent {\bf Procedure} {\it Main}: \\
Every set in $\mathfrak{M}$ is empty; \\
{\bf for each} pixel $p(i, j) = t$:
\begin{itemize}
\item[] {\bf if} there exists a fixed size rule $A \to t \in R$, \\
{\bf then} put $A$ into the set $\mathfrak{M}(i, j; i, j)$;
\end{itemize}

% {\bf remark} the next two loops consider every size $(v,h) \in \{1,\ldots,m\} \times \{1,\ldots,n\}$
% \\
\noindent
{\bf for each} $(v, h) \in \{ 1, \ldots, m \} \times \{ 1, \ldots, n \}$:
%\noindent {\bf for each} $h$, with $1 \le h \le n$:
\begin{itemize}
\item[] {\bf for each} $(i,j) \in  \{1,\ldots,m-v\} \times \{1,\ldots,n-h\}$:
\begin{itemize}
\item[] $\mathfrak{D}$ := Compute$\mathfrak{D}(\mathfrak{M}, (i, j; k, l))$;
\item[] {\bf for each} variable size rule $(A \to \omega) \in R$:
\begin{itemize}
\item[] {\bf if} {\it CheckRule}($\mathfrak{D}, \omega, (i,j; i+v-1,j+h-1)$), \\
{\bf then} put $A$ into the set $\mathfrak{M}(i,j; i+v-1,j+h-1)$;
\end{itemize}
\end{itemize}
\end{itemize}
%{\bf if} $S \in \mathfrak{M}(1,1; m,n)$ {\bf then return} $True$ {\bf else return} $False$. 
{\bf return} $\mathfrak{M}$.
\caption{Main}\label{main}
\end{figure}

\subsubsection{Correctness and complexity of parsing}

We start with a technical lemma, used to prove the correctness of the
CheckRule procedure.

\begin{lemma}\label{L1}
  Let $\omega$ be a regional set of tiles and $d$ a
  subdomain. CheckRule($\omega$, $d$) returns true iff there exists a
  rule $C \to \omega$, such that $(p_0, \pi_0) \Rightarrow_G (p_1,
  \pi_1)$, where $d \in \pi_0$, and $\mathrm{spic}(p_0,d)$ is a
  $C$-picture.
\end{lemma}

\begin{pf}
  By construction, a true output of CheckRule($\omega$, $d$) is
  equivalent to the fact that there exist
$q \in LOC(\omega)$ and
 a partition of $d$ into 
the subdomains $d_1, d_2, \ldots, d_r$, such that:
\begin{enumerate}
\item every $\mathrm{spic}(q,d_j)$ is an
  $A$-picture, for some nonterminal $A \in \mathfrak{M}(d_j)$; 
\item
  if $\mathrm{spic}(q,d_j)$ is an $A$-picture, then for no
  $d_k \ne d_j$ the subpicture $\mathrm{spic}(q,d_k)$ is an $A$-picture.
\end{enumerate}
This means that $\Pi(q) \oplus d$
is the HR partition $\{d_1, d_2, \ldots, d_r\}$.
Moreover, starting from $(p_0,\pi_0)$, where
$\mathrm{spic}(p_0,d)$ is a $C$-picture, it is possible to apply a
rule $C \to \omega$ in a derivation step $(p_0, \pi_0) \Rightarrow_G
(p_1, \pi_1)$, where $\pi_0 = \{d, d'_1, d'_2, \ldots, d'_n\}$, $\pi_1
= $  $\{d'_1, d'_2,$ $ \ldots, d'_n\} \cup $ $\{d_1, d_2, \ldots, d_r\}$,
and $q = \mathrm{spic}(p_1,d) \in LOC(\omega)$. \qed
\end{pf}

After this, the correctness is easy to prove,
analogously to the 1D case \cite{CKY}.

\begin{theorem}
  $\mathfrak{M}(d) = \{ A \in N \mid A
  \stackrel{*}{\Rightarrow}_G \mathrm{spic}(p,d) \}$.
\end{theorem}

\begin{pf}
The proof is by induction on derivation steps.

{\em Base}: $d = (i,j;i,j)$. This means that 
  $|\mathrm{spic}(p,d)| = (1,1)$.
  Hence, $A \stackrel{*}{\Rightarrow}_G \mathrm{spic}(p,d)$ iff $A \to \mathrm{spic}(p,d) \in R$.
This case is handled by the first loop of procedure Main, the one over each pixel $p(i,j)$. 
If $ \mathrm{spic}(p,d) = t$, and there exists a rule $A \to t$, then the algorithm puts $A$ into
$\mathfrak{M}(d)$.
Vice versa, $A \in \mathfrak{M}(d)$ means that the algorithm has put
$A$ in the set, therefore there must exist a rule $A \to \mathrm{spic}(p,d)$.

{\em Induction}: let us consider 
$d = (i,j;i+v-1,j+h-1)$, $v > 1$, or
$h > 1$, or
both.
We prove that $A \stackrel{*}{\Rightarrow}_G \mathrm{spic}(p,d)$
implies $A \in \mathfrak{M}(d)$.
In this case, the size of the subpicture is not $(1,1)$, therefore
the first rule used in the derivation
$A \stackrel{*}{\Rightarrow}_G \mathrm{spic}(p,d)$
is a variable size rule $A \to \omega$.
Thanks to the two nested loops with control variables $(v,h)$ and $(i,j)$, when the
algorithm considers $d$, it has already considered all its subdomains $d_1, d_2, \ldots, d_k$.
By the induction hypothesis, for every $1 \le j \le k$, 
$B \stackrel{*}{\Rightarrow}_G \mathrm{spic}(p,d_j)$ implies $B \in \mathfrak{M}(d_j)$.
Hence (Lemma \ref{L1}), CheckRule($\omega,d$) must be true, and the algorithm puts $A$ in $\mathfrak{M}(d)$.

Next, we prove that
$A \in \mathfrak{M}(d)$ implies
$A \stackrel{*}{\Rightarrow}_G \mathrm{spic}(p,d)$.
$A \in \mathfrak{M}(d)$ means that procedure Main
has put $A$ in the set. 
Therefore, CheckRule($\omega,d$) must be true. 
Thanks to Lemma \ref{L1}, this is equivalent to the 
existence of an applicable variable size rule $A \to \omega$ for the first 
step of the derivation $A \stackrel{*}{\Rightarrow}_G \mathrm{spic}(p,d)$.
The rest of the derivation holds by induction hypothesis.
\qed 
\end{pf}

\begin{theorem}\label{TheorPolynomiality}
 The parsing problem for $\LL(RTG)$ has temporal complexity that is
 polynomial with respect to the input picture size. 
\end{theorem}

\begin{pf}
  First, it is straightforward to see that
  {\em Compute$\mathfrak{D}$} performs a number of operations that is 
$O\left(|N| \cdot m^2 n^2\right)$. 

Let us now consider the {\em CheckRule} procedure.  This procedure performs a
loop for each element of the subdomains vector, which contains a number of
elements that is less than $(m^2 n^2)^{|N|}$, 
  and nested loops on $\mathcal{H}_\omega$
  and $\mathcal{V}_\omega$. Therefore the number of check performed by it
  is dominated by a value that is 
  \[
O\left(
   (m^2 n^2)^{|N|} \cdot 
   \underset{A \to \omega \in R}{\mathrm{max}}\{|\mathcal{H}_\omega|,
   |\mathcal{V}_\omega|\}
\right).
\]

%After computing $\mathfrak{D}$,

Coming finally to the {\em Main} procedure, we note that its core part
consists of two nested loops, over two sets that are at most $m \cdot n$ each.
The body of these two loops consists in a call to $\mathrm{Compute}\mathfrak{D}$, and
then another loop over the grammar rules, comprising a call to {\em
CheckRule} (hence the dominant part).

Therefore, the number of operations performed is at most 
\[
O\left(
    |R| \cdot \underset{A \to \omega \in
     R}{\mathrm{max}}\{|\mathcal{H}_\omega|, |\mathcal{V}_\omega|\}
   \cdot
   (m^2 n^2)^{|N|} \cdot m^2 n^2
\right).
\]
Each of these operations can be done in polynomial time in every
reasonable machine model, therefore the resulting time complexity is 
polynomial w.r.t. the picture size.  
\qed
\end{pf}

The property of having polynomial time complexity for picture
recognition, united with the rather simple and intuitively pleasing
form of RTG rules, should make them a worth addition to the
series of array rewriting grammar models conceived in past years.

%\subsubsection{Comparison with other language families}\label{SectionComparisons}

%%% Local Variables: 
%%% mode: latex
%%% TeX-master: "grammars-arxiv"
%%% End: 

\section{Comparison with other language families}\label{SectionComparisons}

In this section we prove or recall some inclusion relations between
grammar models and corresponding language families. To this end we
rely on the examples of Section \ref{SectionRegionalGrammars}, and on
the separation of complexity classes.

We start by comparing regional tile grammars and tiling systems.
To this end, we adapt a proof and an example introduced by \prusa{} in
\cite{Prusa2004}.

\begin{example}\label{ltex}
Consider a language $L_{lt}$ over the alphabet $\Sigma = \{0,0',1,1',x,x'\}$
where the ``primed'' symbols are used on the diagonal.
A picture $p$ is in $L_{lt}$ if, and only if:

\begin{enumerate}
\item $p$ is a square picture of odd size;

\item $p(i,j) \in  \{0,1,x\}$, when $i \ne j$;  $p(i,j) \in  \{0',1',x'\}$,
otherwise.

\item $p(i,j) \in \{x,x'\}$ iff $i$ and $j$ are odd;

\item if $p(i,j) \in \{1,1'\}$ then the $i$-th row or the $j$-th
column (or both) is made of symbols taken from $\{1, 1'\}$.
\end{enumerate}

\noindent An example picture is shown in Figure~\ref{ltex:fig}. Primed symbols
by definition appear only on the main diagonal, and are used to have only square
pictures.  
It is quite easy to see that $L_{lt}$ is a locally testable language, definable
through a set of 3-tiles.

\end{example}

\begin{figure}
\[
\begin{array}{|ccccccc|}
\hline
x'& 1 & x & 1 & x & 0 & x\\
0 & 1'& 0 & 1 & 0 & 0 & 0\\
x & 1 & x'& 1 & x & 0 & x\\
1 & 1 & 1 & 1'& 1 & 1 & 1\\
x & 1 & x & 1 & x'& 0 & x\\
0 & 1 & 0 & 1 & 0 & 0'& 0\\
x & 1 & x & 1 & x & 0 & x'\\
\hline
  \end{array}
\]
\caption{A picture of the language $L_{lt}$ of Example~\ref{ltex}}\label{ltex:fig}
\end{figure}

\begin{proposition}\label{lt-vs-rtg}
$\LL(RTG)$  and  $\LL(LT)$ are incomparable.
\end{proposition}

\begin{pf}
First, we know from~\cite{GR-two-dl} that $\LL(LT) \subset
\LL(TS)$, and that the non-TS language of
palindromes, used in \cite{TRG1} to prove that tiling systems are
strictly included in tile grammars, is also a RTG language, obtained
by a $90^{\text{o}}$ rotation of Example \ref{PalindromicRowsEx}.

To end the proof, we need a language that is in $\LL(LT)$ but not in
$\LL(RTG)$.
Let $G=(\Sigma, N, S, R)$ be a RTG such that $L(G) = L_{lt}$ of
Example~\ref{ltex}. W.l.o.g., we assume that $R$ does not contain chain rules.
We consider a natural number $n = 2k+1$ big enough to comply with the
requirements presented in the rest of the proof. 

First, let $L_1$ be $\{ p \in L_{lt} \mid |p| = (n,n) \}$. Clearly, $|L_1| =
2^{n-1}$, and it contains at least $\lceil 2^{n-1}/|R| \rceil$ pictures that can be generated
in the first step by the same rule.

We now fix a rule, e.g. $S \to \alpha$, and let $L_2$ be the subset of
$L_1$ generated by this rule. 
In a $n$ by $n$ picture, the number of possible partitions
in homogeneous subpictures is less than $(n^4)^{|N|}$. This means that there exists a set
$L_3 \subseteq L_2$, having size $|L_3| \ge {\frac{2^{n-1}}{|R|\cdot
n^{4|N|}}}$ such that every picture in it was generated by $G$ starting
with the same rule $S \to \alpha$, and such that the initial $S$-homogeneous
picture was replaced by the same $s \in LOC(\alpha)$. 

Depending on the chosen rule's right part, i.e. $\alpha$, we now identify a row or
a column of the picture in an odd position, and call it $\lambda$. We have two
cases: either (1) every $s \in LOC(\alpha)$ is made of homogeneous subpictures
having all both width and height less than $n$; or (2) in every $s \in
LOC(\alpha)$ there is at least one homogeneous subpicture $s'$ having width or
height equal to $n$ (but clearly not both, because we are not considering chain
rules). In case (1), let $\lambda$ be the first row. In case (2), let $\lambda$
be one of the rows or columns in an odd position and completely contained in
$s'$.   

Let $L_4$ be a subset of $L_2$ such that every picture in it has the same
$\lambda$. Because of its definition, if we fix an odd row of pictures in
$L_{lt}$, then columns of even indexes that are completely filled by $1$ and
$1'$ are determined by it (if we fix an odd column, it is analogous but with
rows). Hence, $|L_4| \le 2^{\frac{n-1}{2}}$.

We can assume that $n$ is sufficiently large so that $|L_3| > |L_4|$,
i.e. there is at least a picture in $L_3$ which is not present in $L_4$. So we are
able to find in $L_3$ two pictures $p$ and $q$ that are generated by the
same initial rule, $S \to \alpha$, with the same initial strong
homogeneous partition (the one determined by $s$), and such
that $\lambda$ in $p$ is different from $\lambda$ in $q$. 
Now consider all the subpictures of $p$ and $q$ that are in the positions
corresponding to the initial strong homogeneous partition. 
Of these subpictures, we consider only the sets $P' = \{p'_1, p'_2, \ldots,
p'_i\}$, and
$Q' = \{q'_1, q'_2, \ldots, q'_j\}$, with
$i,j \le |N|$, that contain $\lambda$ in $p$ and in $q$, respectively.
If we replace in $p_1$ all the elements of $P'$ with the elements in $Q'$, we
obtain a picture which is derivable from $S \to \alpha$, but it is not in
$L_{lt}$, because it contains columns (or rows in some cases (2)) that are not
compatible with the fixed $\lambda$. \qed 
\end{pf}
%   On one hand, it is easy to see that the non-TS language of
%   palindromes, used in \cite{TRG1} to prove that tiling systems are
%   strictly included in tile grammars, is also a RTG language, obtained
%   by a $90^{\text{o}}$ rotation of Example \ref{PalindromicRowsEx}.
%   On the other hand, we know that parsing tiling systems is
%   NP-complete \cite{lindgren}, and parsing $\LL(RTG)$ has polynomial
%   time complexity (Proposition \ref{TheorPolynomiality}). \qed

The fact that $\LL(LT) \subset \LL(TS)$ implies the following
statement.

\begin{corollary}
 $\LL(RTG)$  and  $\LL(TS)$ are incomparable.
\end{corollary}
 
This last result, together with the facts that RTG rules are a
restricted form of TG rules, and that $\LL(TS) \subset \LL(TG)$, gives us the
following:

\begin{corollary}
$\LL(RTG) \subset \LL(TG)$.
\end{corollary}

% \begin{pf}
%   We have seen in Sect. \ref{SectionRegionalGrammars} that RTG rules
%   are a restricted form of TG rules, characterized by the constraint
%   of regional tiling.  To show that inclusion is strict, we observe
%   that the picture recognition problem for tile grammars is
%   NP-complete (Proposition \ref{parse-tg-is-np}). \qed
% \end{pf}

\subsection{Context-free Kolam grammars}\label{sec:kolam}

This class of grammars has been introduced by Siromoney et al. 
\cite{GSiromoney-RSiromoney-KKrithivasan:73b} under the name ``Array
grammars'', later renamed ``Kolam Array grammars'' in order to avoid
confusion with Rosenfeld's homonymous model. 
Much later Matz reinvented the same model \cite{STACS::Matz1997}
(considering only CF rules).
We prefer to keep the historical name, CF
Kolam grammars (CFKG), and to use the more succint definition of Matz.

\begin{definition}
    A {\em sentential form} over an alphabet $V$ is a non-empty well-par\-enthe\-sized
    expression using the two concatenation operators, $\ominus$ and
    $\obar$, and symbols taken from $V$.
    $\mathcal{SF}(V)$ denotes the set of all
    sentential forms over $V$.
    A sentential form $\phi$ defines either one picture over $V$
    denoted by $\llparenthesis \phi \rrparenthesis$, or none.
\end{definition}

For example,
$\phi_1 = \left((a \obar b) \ominus (b \obar a)\right) \in
\mathcal{SF}(\{a,b\})$
and
$\llparenthesis \phi_1 \rrparenthesis$ is the picture
$\begin{array}{|cc|}
\hline
a & b \\
b & a \\
\hline
\end{array}$.
On the other hand $\phi_2 = \left((a \obar b) \ominus a\right)$ denotes no
picture, since the two arguments of the $\ominus$ operator have different column numbers.

CF Kolam grammars are defined analogously to CF string grammars. Derivation is
similar: a sentential form over terminal and nonterminal symbols
results from the preceding one by replacing a
nonterminal with some corresponding right hand side of a rule.
The end of a derivation is reached when the
sentential form does not contain any nonterminal symbols. If this resulting
form denotes a picture, then that picture is generated by the grammar.

\begin{definition}\label{def:kolam}
    A {\em context-free Kolam grammar (CFKG)} is a tuple $G =
    (\Sigma,N,S,R)$, where $\Sigma$ is the finite set of {\em terminal}
    symbols, disjoint from the set $N$ of {\em nonterminal} symbols;
    $S \in N$ is the {\em start} symbol; and $R \subseteq N \times \mathcal{SF}(N
    \cup \Sigma)$ is the set of {\em rules}. A rule $(A,\phi) \in
    R$
    will be written as
    $A \to \phi$.
\end{definition}

For a grammar $G$, we define the {\em derivation} relation $\Rightarrow_G$ on
the sentential forms $\mathcal{SF}(N \cup \Sigma)$ by $\psi_1 \Rightarrow_G \psi_2$
iff there is
some rule $A \to \phi$, such that $\psi_2$ results from $\psi_1$ by
replacing an occurrence of $A$ by $\phi$. As usual,
$\stackrel{*}{\Rightarrow}_G$ denotes the reflexive and transitive closure.
Notice that the derivation thus defined rewrites strings, not pictures.

From the derived sentential form, one then obtains the denoted
picture. The picture language generated by $G$ is the set
\[
L(G) = \{ \llparenthesis \psi \rrparenthesis \mid \psi \in
\mathcal{SF}(\Sigma), S \stackrel{*}{\Rightarrow}_G \psi \}.
\]
With a slight abuse of notation, we will often write $A
\stackrel{*}{\Rightarrow}_G p$, with $A \in N, p \in \Sigma^{++}$, instead of
$\exists \phi : A \stackrel{*}{\Rightarrow}_G \phi,  \llparenthesis \phi
\rrparenthesis = p$.

It is convenient to consider a normal form with exactly two or zero nonterminals in the right part of a rule
\cite{STACS::Matz1997}.

\begin{definition}\label{def:cnf}
    A CF Kolam grammar $G = (\Sigma,N,S,R)$,
    is in {\em Chomsky Normal Form} iff every rule in $R$ has the form
    either $A \to t$, or $A \to B \ominus C$, or $A \to B
    \obar C$, where $A,B,C \in N$, and $t \in \Sigma$.
\end{definition}

We know from \cite{STACS::Matz1997} that for every CFKG $G$, if $L(G)$ does
not contain the empty picture,
there exists a
CFKG $G'$ in Chomsky Normal Form, such that $L(G) = L(G')$.
Also, the classical algorithm to translate a
string grammar into Chomsky Normal Form can be easily adapted to CFKGs.

\begin{example}\label{ex:kolamgrammar}

    The following Chomsky Normal Form grammar $G_5$
 % of Figure \ref{fig:ex:kolamgrammar} 
    defines the set of pictures such that each column is an odd length
    palindrome.
%
%\begin{figure}[h!]
\[
\begin{array}{ccl}  
    S & \to & V \obar S \mid A_1 \ominus A_2 \mid B_1 \ominus B_2 \mid a \mid b \\
    V & \to &  A_1 \ominus A_2 \mid B_1 \ominus B_2 \mid a \mid b \\
    A_2 & \to & V \ominus A_1 \mid a  \\
    B_2 & \to & V \ominus B_1 \mid b  \\
    A_1 & \to & a \\
    B_1 & \to & b. \\
  \end{array}
\]
%\caption{The CFKG $G_5$ of Example \ref{ex:kolamgrammar}.}\label{fig:ex:kolamgrammar}
%\end{figure}

\end{example}

\subsubsection{Comparison with other models}

First, we sketchily and intuitively show that the original CF Kolam definition is
equivalent to the one introduced by Matz.
The following description is directly taken from
\cite{GSiromoney-RSiromoney-KKrithivasan:73b}.

\smallskip
{\it
	Let $G = (\Sigma,N,S,R)$, be a {\em Kolam context-free grammar}, where
	$N = N_1 \cup N_2$, $N_1$ a finite set of {\em nonterminals}, $N_2$
	a finite set of {\em intermediates}, $\Sigma$ a finite set of {\em
	terminals}, $R = R_1 \cup R_2 \cup R_3$, $R_1$ a finite set of
	{\em nonterminal rules}, $R_2$ a finite set of {\em
	intermediate rules},
	$R_3$ a finite set of {\em terminal rules}. $S \in N_1$ is the
	start symbol.

	\noindent
        $R_1$ is a set of pairs $(A,B)$ (written $A \to B$),
	$A \in N_1$, $B \in (N_1 \cup N_2)^{+\obar}$ or $B \in
	(N_1 \cup N_2)^{+\ominus}$.

        \noindent
	$R_2$ is a set of pairs $(B,C)$, $B \in N_2$, $C \in (N_2 \cup \{x_1, x_2,
	\cdots, x_k\})^{+\obar}$, \\
        with $x_1,\cdots,x_k \in \Sigma^{++}$,
	$|x_i|_{row} = |x_{i+1}|_{row}$, $1 \le i < k$; \\
	or  $C \in (N_2 \cup \{x_1, x_2, \cdots, x_k\})^{+\ominus}$,
	with $x_1,\cdots,x_k \in \Sigma^{++}$,
	$|x_i|_{col} = |x_{i+1}|_{col}$, $1 \le i < k$.

        \noindent
	$R_3$ is a set of pairs $(A,t)$, $A \in (N_1 \cup N_2)$ and $t \in
	\Sigma^{++}$.

        \medskip

        \noindent
	(Derivation) If $A$ is an intermediate, then the intermediate
	language generated by $A$ is 
	$
	M_A = \{x \mid A
	\stackrel{*}{\Rightarrow} x, x \in \{x_1, \cdots, x_k\}^{+\obar},
	x_j \in \Sigma^{++}, |x_i|_{row} = |x_{i+1}|_{row}, 
	1 \le i < k\}
	$ 
	or
	$
	M_A = \{x \mid A
	\stackrel{*}{\Rightarrow} x, x \in \{x_1, \cdots, x_k\}^{+\ominus},
	x_j \in \Sigma^{++}, |x_i|_{col} = |x_{i+1}|_{col}, 
	1 \le i < k\}.
	$
	Derivation proceeds as follows. Starting from $S$, nonterminal
	rules are applied without any restriction as in a string grammar,
	till all the nonterminals are replaced, introducing parentheses
	whenever necessary. Now replace for each intermediate $A$ in $N_2$
	elements from $M_A$, subject to the conditions imposed by $\obar$,
	$\ominus$. The replacements start from the innermost parenthesis
	and proceeds outwards. The derivation comes to an end if the
	condition for $\ominus$ or $\obar$ is not satisfied.
}% end \it
\smallskip

Grammar $G_5$ of Example \ref{ex:kolamgrammar} complies
with this definition. In it, $A_1$ and $B_1$ are intermediates.

It is very easy to see that the original definition of CF Kolam
grammars is equivalent to the new one given by Matz. Right part of
rules are made of vertical or horizontal concatenations of
nonterminals or fixed terminal pictures.
So we can define an equivalent grammar
that is as stated in Definition~\ref{def:kolam}, by translating the right
part of rules that contain terminal pictures $x_1, x_2, \ldots, x_p$, decomposing each
picture $x_i$ in a
sentential form $\phi$ such that $x_i = \llparenthesis \phi
\rrparenthesis$. Vertical or horizontal concatenations are then
treated analogously (e.g. we translate $A B$ into $(A \obar
B)$). Clearly, we do not need to distinguish nonterminals from
intermediate symbols.

% The fact that the grammars defined in Definition \ref{def:kolam} and those
% defined in Definition \ref{def:kolam-old} are equivalent is a
% straightforward consequence of the fact that both can define any finite 2D
% language, and are both closed w.r.t. $\obar$,$\ominus$,$ ^{+\obar}$,$
% ^{+\ominus}$.

\begin{proposition}\label{kolam-in-rtg}
$\LL(CFKG) \subset \LL(RTG)$.
%  The family of CF Kolam array grammar
%  \cite{GSiromoney-RSiromoney-KKrithivasan:73b} (i.e. also
%  \cite{STACS::Matz1997}) languages is strictly included in the family
%  of RTG languages.
\end{proposition}

\begin{pf}
  In \cite{TRG1} a construction is given to prove that a CF Kolam
  grammar (in the form defined by Matz \cite{STACS::Matz1997}) can be
  transformed into a TG. It turns out that the TG thus constructed is
  a RTG.

  Sketchily, consider a CF Kolam grammar $G$ in CNF. Rules $A \to t, t \in
  \Sigma$ are identical in the two models and generate the same kind of
  languages (i.e. single terminal symbols).
  Rules $A \to B \obar C$ of $G$ are equivalent to RTG rules having
  the following form: 
  \[
  A \to
  \left\llbracket
  \begin{array}{cccccc}
	  \# & \# & \# & \# & \# & \# \\
	  \# & B & B & C & C & \# \\
	  \# & B & B & C & C & \# \\
	  \# & \# & \# & \# & \# & \# \\
  \end{array}
  \right\rrbracket
  \]

  Rules $A \to B \ominus C$ of $G$ are equivalent to RTG rules having
  the following form: 
  \[
  A \to
  \left\llbracket
  \begin{array}{cccccc}
	  \#  & \# & \# & \# \\
	  \# & B & B & \# \\
	  \# & B & B & \# \\
	  \# & C & C & \# \\
	  \# & C & C & \# \\
	  \#  & \# & \# & \# \\
  \end{array}
  \right\rrbracket
  \]
  
  The inclusion is strict, because the language of Example
  \ref{twoBytwoDividerEx} was shown by Matz \cite{STACS::Matz1997} to
  trespass the generative capacity of his grammars. \qed
\end{pf}

The fact that the picture recognition problem for CF Kolam
grammars has been recently proved \cite{KolamParsing} to be polynomial
in time of course follows from the above inclusion property and from
Theorem \ref{TheorPolynomiality}.

For the special case of CF Kolam grammars in Chomsky Normal form (CNF), we
note that the parsing time complexity is $O(m^2 n^2 (m+n))$
\cite{KolamParsing}.  Some of the reasons of this significant difference
are the following.  Kolam grammars in CNF are much simpler, because in the
right part of a rule there are at most two distinct nonterminals.  So,
checking if a rule is applicable has complexity which is linear with
respect to the picture width or height.

%%% Local Variables: 
%%% mode: latex
%%% TeX-master: "paper"
%%% End: 

\subsection{\prusa{}'s context-free grammars}\label{sec:prusa}

In the quest for generality, D. \prusa{}  \cite{Prusa2004} has
recently defined a grammar model that extends CF Kolam
rules, gaining some generative capacity.  The model is for instance
able to generate the language of Example \ref{twoBytwoDividerEx}.

\subsubsection{Definitions}

The following definitions are taken and adapted from
\cite{Prusa2001,Prusa2004}.

\begin{definition}
	A 2D CF \prusa{} grammar (PG) is a tuple $(\Sigma,N,S,R)$, where
	$\Sigma$ is the finite set of {\em terminal} symbols, disjoint from
	the set $N$ of {\em nonterminal} symbols; $S \in N$ is the {\em
	start} symbol; and $R \subseteq N \times (N \cup
	\Sigma)^{++}$ is the set of {\em rules}. 
\end{definition}

\begin{definition}
	Let $G = (\Sigma,N,S,R)$ be a PG. We define a picture language
	$L(G,A)$ over $\Sigma$ for every $A \in N$.
	The definition is given by the following recursive descriptions:

	\begin{itemize}
		\item[(i)] If $A \to w$ is in $R$, and $w \in \Sigma^{++}$, then
			$w \in L(G,A)$.
		\item[(ii)] Let $A \to w$ be a production in $R$, $w =
			(N\cup\Sigma)^{(m,n)}$, for some $m,n \ge 1$. Let $p_{i,j}$, with $1 \le i \le m$, $1 \le
			j \le n$, be pictures such that:
			
			\begin{enumerate}
				\item 
			 if $w(i,j) \in
			\Sigma$, then $p_{i,j} = w(i,j)$; 
		\item if $w(i,j) \in
			N$, then $p_{i,j} \in L(G,w(i,j))$; 
		\item let $P_k =
			p_{k,1} \obar p_{k,2} \obar \cdots \obar p_{k,n}$.
			For any $1 \le i < m$, $1 \le j \le n$, 
			$|p_{i,j}|_{col} = |p_{i+1,j}|_{col}$; and  
			$P = P_1 \ominus P_2 \ominus \cdots \ominus P_m.$ 
			\end{enumerate}
			Then $P \in L(G,A)$. 
	\end{itemize}

	The set $L(G,A)$ contains all and only the pictures that can be
	obtained by applying a finite sequence of rules (i) and (ii). The
	language $L(G)$ generated by grammar $G$ is defined as the
	language $L(G,S)$.
\end{definition}

Informally, rules can either be terminal rules, in this case managed
exactly as tile grammars or Kolam grammars, or have a picture as right
part. In this latter case, the right part is seen as a ``grid'', where
nonterminals can be replaced by other pictures, but maintaining its
grid-like structure. Note that the grid meshes may differ in size.

\begin{example}\label{prusa-example}
The grammar $G_6$ of Figure \ref{prusa-grammar} generates the language of pictures with one row and
one column of $b$'s in a background of $a$'s (see Example
\ref{twoBytwoDividerEx}).

\begin{figure}[h!]
\[ 
S \to 
\begin{array}{ccc}
	A & V & A \\
	H & b & H \\
	A & V & A \\
\end{array},
\ \ 
A \to A M \mid M,
\ \ 
M \to 
\begin{array}{c}
	a \\
	M \\
\end{array} \mid a,
\]
\[ 
V \to 
\begin{array}{c}
	b \\
	V \\
\end{array} \mid b,
\ \
H \to b H \mid b.
\]
\caption{PG $G_6$ of Example
  \ref{prusa-example}.}\label{prusa-grammar}
\end{figure}

\end{example}

We now introduce a normal form for \prusa{} grammars:

\begin{definition}\label{def:prusa:cnf}
	A \prusa{} grammar $G = (\Sigma,N,S,R)$,
	is in {\em Nonterminal Normal Form} (NNF) iff every rule in $R$ has the form
	either $A \to t$, or $A \to w$, where $A \in N$, $w \in N^{++}$, 
	and $t \in \Sigma$.
\end{definition}

\subsubsection{Comparison with other models}

To compare \prusa{} grammars with tile grammars, we note that the
two models are different in their derivations. Tile grammars start from a
picture made of $S$'s having a fixed size, and being every derivation step
isometric, the resulting picture, if any, has the same size.
On the other hand, \prusa{} grammars start from a single $S$ symbol, and then ``grow''
the picture derivation step by derivation step, obtaining, if any, a
usually larger picture.

First, we prove that the language
of Example~\ref{MisalignedPalindromicRowsEx} cannot be defined by
\prusa{} grammars, so the language families of regional tile grammars
and \prusa{} grammars are different. To this aim, we use a technique
analogous to the one introduced for proving
Proposition~\ref{lt-vs-rtg}.

\begin{proposition}
$\LL(PG) \ne \LL(RTG)$.
\end{proposition}
\begin{pf}
Let $G = (\Sigma,N,S,R)$ be a PG such that $L(G) = L(G_4)$,
where $G_4$ is the RTG presented
in Example~\ref{MisalignedPalindromicRowsEx}.
W.l.o.g. we assume that $R$ does not contain chain rules, and consider
a natural number $n$ big enough to comply with the requirements
of the rest of the proof.
First we consider $L_0 \subset L(G_4)$, on alphabet $\{a, c\}$, where the
palindromes are made exclusively of $a$ symbols. Suppose that pictures in
$L_0$ are generated by a rule 
$S \to \begin{smallmatrix} A \\ B\end{smallmatrix}$, $A,B \in N$. In this case
it is easy to see that $A$ must generate strings $a^i c^j$, with $i+j = n$,
while $B$ generates strings $c^k a^l$, $k+l = n$. But it is possible to take $i
< k$, thus obtaining pictures that are not in $L(G_4)$. So we can assume that
the starting rules are like $S \to w$, with $w$ having at most two rows and at
least two columns.

Now consider $L_1 \subset L(G_4)$, in which every picture has two
rows, $3n$ columns, and is such that the two $c$-homogeneous subpictures in it
have size $(1,n)$; hence $|L_1| = 2^{2n}$. The set $L_1$ contains at least  
$\lceil 2^{2n}/|R| \rceil$ pictures that can be generated
in the first step by the same rule.

We fix a rule, e.g. $S \to w$, with $|w| = (a,b)$, $1 \le a \le 2$, $b > 1$, and let
$L_2$ be the subset of $L_1$ generated by this rule. W.l.o.g. we assume that $n
> b$, so each nonterminal in $w$ generates a subpicture (that in the rest of the
proof we will index by
$p_{i,j}$, $1 \le i \le a$, $1 \le j \le b$) having at most two rows and one
column. Being the number of different sequences $|p_{1,1}|_{col}$,
$|p_{1,2}|_{col}$, \ldots $|p_{1,b}|_{col}$, $|p_{1,1}|_{row}$, $|p_{2,1}|_{row}$
limited by $2(3n)^{b}$ (each $|p_{1,i}|_{col}$ is less than $3n$ and at most there
are two rows), there exists a subset $L_3$ of $L_2$, having size $|L_3|
\ge {2^{2n}}/\left(2|R|(3n)^{b}\right)$, in which for any two pictures $p$ and $p'$,
and for every $i,j$, $|p_{i,j}|$ is equal to $p'_{i,j}$. 

Let $L_4$ be a subset of $L_2$ such that every picture in it is like
$\begin{smallmatrix}
q^R & q  & c^n\\
c^n & q  & q^R\\
     \end{smallmatrix}$,
(i.e. the central third of the picture is made
of two equal rows). Clearly, $|L_4| \le 2^n$. 

We can assume that $n$ is large enough so that $|L_3| > |L_4|$. But this means
that in $L_3$ there are two different pictures 
$p = \begin{smallmatrix}
q^R & q  & c^n\\
c^n & s  & s^R\\
     \end{smallmatrix}$,
and
$p' = \begin{smallmatrix}
q'^R & q'  & c^n\\
c^n & s'  & s'^R\\
     \end{smallmatrix}$, with $q \ne s$, $q' \ne s'$, and
(1) $q \ne q'$ or (2) $s \ne s'$. We know
that $b>1$, so if we replace $p_{1,1}$ and $p_{2,1}$ (if $a=2$) in $p$ with
$p'_{1,1}$ and $p'_{2,1}$, in case (1), we obtain a picture generated by $G$
that is not in $L(G_4)$. Case (2) is analogous, but considers the right part of
$p$, i.e. $p_{1,b}$ and $p_{2,b}$.
\qed
\end{pf}
  
Indeed, \prusa{} grammars can be seen as a restricted form of regional
tile grammars, as stated by the following proposition.

\begin{proposition}\label{prusa-in-rtg}
$\LL(PG) \subset \LL(RTG)$.
\end{proposition}
\begin{pf}

  Consider a PG in NNF $G$. First of all, we assume without loss of
  generality that for any rule, nonterminals used in its right part
  are all different. If this is not the case, e.g. assume that we have
  a rule 
	\[
	A \to
\begin{array}{cc}
 X & Y  \\
 Z & X  \\
\end{array},
\]
then we can rename one of the $X$ symbols to a freshly introduced
nonterminal $X'$, and then add the chain rule $X' \to X$.

Let us define a RTG $G'$ equivalent to
	$G$.
	Terminal rules are easily treated.
	For a nonterminal rule of $G$,  e.g. 
	\[
	A \to
\begin{array}{ccc}
 B_{1,1} & \ldots & B_{1,k}  \\
 \vdots & \ddots & \vdots  \\
 B_{h,1} & \ldots & B_{h,k}  \\
\end{array}
\]
we introduce the following rule in $G'$:
	\[
	A \to
	\left\llbracket
\begin{array}{ccccccc}
\# & \# & \# & \ldots & \# & \# & \# \\
\# & B_{1,1} &  B_{1,1} & \ldots & B_{1,k} & B_{1,k} & \# \\
\# & B_{1,1} &  B_{1,1} & \ldots & B_{1,k} & B_{1,k} & \# \\
\vdots & \vdots & \vdots & \ddots & \vdots & \vdots & \vdots \\
\# & B_{h,1} & B_{h,1} & \ldots & B_{h,k} & B_{h,k} & \# \\
\# & B_{h,1} & B_{h,1} & \ldots & B_{h,k} & B_{h,k} & \# \\
\# & \# & \# & \ldots & \# & \# & \# \\
\end{array}
\right\rrbracket.
\]
Note that each nonterminal $B_{i,j}$ is repeated four times in the
right part of the rule, so to have the tile
$
\begin{array}{|cc|}
\hline
B_{i,j} & B_{i,j} \\
B_{i,j} & B_{i,j} \\
\hline
\end{array}
$,
that can be used to ``cover'' a rectangular area of any size.

% where the nonterminal symbols $B'_{i,j}$ are chosen to avoid repetition
% (because of the regional constraint), i.e. $B'_{i,j} \ne B'_{i',j'}$, for
% any $i,j$,$i',j'$.
% This means that $B'_{i,j}$ may
% be either equal to $B_{i,j}$, or freshly introduced. In this latter case, we add
% the chain rule 
% \[
% B'_{i,j} \to
% \left\llbracket
% \begin{array}{cccc}
%     \#  & \#   & \#  & \# \\
%     \#  & B_{i,j}  &  B_{i,j}  & \# \\
%     \#  & B_{i,j}  &  B_{i,j}  & \# \\
%     \#  & \#   & \#  & \# \\
%   \end{array}
% \right\rrbracket
% \]

Essentially, \prusa{} grammars can be seen as RTG's with
the additional constraint that tiles used in the right parts of rules
must not have one of these forms:
\[
\begin{array}{|cc|}
\hline
A & B \\
C & C \\
\hline
\end{array}
, \ 
\begin{array}{|cc|}
\hline
A & C \\
B & C \\
\hline
\end{array}
, \ 
\begin{array}{|cc|}
\hline
C & C \\
A & B \\
\hline
\end{array}
, \
\begin{array}{|cc|}
\hline
C & A \\
C & B \\
\hline
\end{array}
\]
with $A,B,C$ all different.
\qed
\end{pf}

\begin{proposition}
$\LL(CFKG) \subset \LL(PG)$.
%The class of CF Kolam grammars languages is properly included in the class of CF
%\prusa{} grammars languages.	
\end{proposition}

\begin{pf}
For containment, it suffices to note that the constraints on tiles of the
corresponding tile grammar, introduced in the proof of Proposition
\ref{prusa-in-rtg}, are a weaker form of the constraints used for proving
Proposition \ref{kolam-in-rtg}. 

The containment is strict, since \prusa{} grammar can generate the language
of one column and one row of $b$'s in a field of $a$'s (see Example
\ref{prusa-example}), while CF Kolam grammar cannot
\cite{STACS::Matz1997}. \qed
\end{pf}

%%% Local Variables: 
%%% mode: latex
%%% TeX-master: "paper"
%%% End: 

\subsection{Grid grammars}\label{sec:grid}

Grid grammars are an interesting formalism defined by Drewes
\cite{Drewes1996},\cite{Drewes2003}. Grid grammars are based on an
extension of quadtrees \cite{finkel1974}, in which the number of
``quadrants'' is not limited to four, but can be $k^2$, with $k \ge 2$
(thus forming a square ``grid'').

Following the tradition of quadtrees, and differently from the other
formalisms presented here, grid grammars generate pictures which are seen
as sets of points on the ``unit square'' delimited by the points (0,0),
(0,1), (1,0), (1,1) of the Cartesian plane.
The following definitions are taken (and partially adapted) from \cite{Drewes2003}.

\smallskip

{\it
Let the unit square be divided by a evenly spaced grid into $k^2$ squares,
for some $k \ge 2$.
A {\em production} of a grid picture grammar consists of a nonterminal
symbol on the left-hand side and
the square grid on the right-hand side, each of the $k^2$ squares in the
grid being either black or white or labelled with a nonterminal.

A {\em derivation} starts with the {\em initial nonterminal} placed in the
unit square. Then productions are applied repeatedly until there is no
nonterminal left, finally yielding a generated picture. As usual, a
production is applied by choosing a square containing a nonterminal $A$ and
a production with left-hand symbol $A$. The nonterminal is then removed
from the square and the square is subdivided into smaller black, white, and
labelled squares according to the right-hand side of the chosen production.
The set of all pictures generated in this manner constitutes the {\em picture
language} generated by the grammar.

A picture generated by a grid picture grammar can be written as a string
expression. Let the unit black square be represented by the symbol $B$, and
the white unit square by $W$. By
definition, each of the remaining pictures in the generated language
consists of $k^2$ subpictures $\pi_{1,1}, \ldots$ $\pi_{1,k}, \ldots$
$\pi_{k,1}, \ldots, \pi_{k,k}$, each scaled by the factor $1/k$, going from
bottom-left $\pi_{1,1}$ to top right $\pi_{k,k}$.
If $t_{i,j}$ is the expression representing $\pi_{i,j}$ (for $1 \le i,j \le
k$), then $[t_{1,1}, \ldots, t_{1,k}, \ldots, t_{k,1}, \ldots, t_{k,k}]$
represents the picture itself (for $k=2$ it is a quadtree).
} % end \it

\medskip

In order to compare such model, in which a picture is in the unit
square and mono-chromatic, with the ones presented in this work, we
introduce a different but basically compatible formalization, in which
the generated pictures are square arrays of symbols, and the terminal
alphabet is not limited to black and white.

\subsubsection{Definitions}

To define grid grammars, we use a technique similar to the one used for
Kolam grammars in Section \ref{sec:kolam}.

\begin{definition}
  For a fixed $k \ge 2$, a {\em sentential form} over an alphabet $V$
  is either a symbol $a \in V$, or $[t_{1,1}, \ldots, t_{1,k}, \ldots,
  t_{k,1}, \ldots, t_{k,k}]$, and every $t_{i,j}$ being a sentential
  form.  $\mathcal{SF}(V)$ denotes the set of all sentential forms
  over $V$.

	A sentential form $\phi$ defines a set of pictures $\llparenthesis \phi
\rrparenthesis$:
\begin{itemize}
	\item $\llparenthesis a \rrparenthesis$, with $a \in V$, represents the set
		$\{a\}^{(n,n)}, n \ge 1$ of all $a$-homogeneous square pictures; 
	\item
		$\llparenthesis [t_{1,1}, \ldots, t_{1,k}, \ldots, t_{k,1},
		\ldots, t_{k,k}] \rrparenthesis$, represents the set of all
		square grid pictures where every $\llparenthesis t_{i,j}
		\rrparenthesis$ has the same size $n \times n$, for $n \ge 1$,
		and $\llparenthesis t_{1,1} \rrparenthesis$
		is at the bottom-left corner, \ldots, $\llparenthesis t_{1,k}
		\rrparenthesis$ is at the bottom right corner, \ldots, and
		$\llparenthesis t_{k,k} \rrparenthesis$ is at the top right corner.
\end{itemize}

\end{definition}

Note that we maintained in the sentential forms the original
convention of starting from the bottom-left position.
For example, consider the sentential form 
\[
\phi = \left[
  [a,b,[a,b,b,a],c], a, B, [b,a,a,b] \right].
\] 
The smallest picture in
$\llparenthesis \phi \rrparenthesis$ is depicted in Figure
\ref{grid:pict}.
\begin{figure}[h!]
\[
\begin{array}{|cccccccc|}
\hline
  B & B & B & B & a & a & b & b \\
  B & B & B & B & a & a & b & b \\
  B & B & B & B & b & b & a & a \\
  B & B & B & B & b & b & a & a \\
  b & a & c & c & a & a & a & a \\
  a & b & c & c & a & a & a & a \\
  a & a & b & b & a & a & a & a \\
  a & a & b & b & a & a & a & a \\
\hline
\end{array}
\]
\caption{Example picture generated by the form $\left[
  [a,b,[a,b,b,a],c], a, B, [b,a,a,b] \right]$.}\label{grid:pict}
\end{figure}

\begin{definition}
    A {\em grid grammar (GG)} is a tuple $G =
    (\Sigma,N,S,R)$, where $\Sigma$ is the finite set of {\em terminal}
    symbols, disjoint from the set $N$ of {\em nonterminal} symbols;
    $S \in N$ is the {\em start} symbol; and $R \subseteq N \times \mathcal{SF}(N
    \cup \Sigma)$ is the set of {\em rules}. A rule $(A,\phi) \in
    R$
    will be written as
    $A \to \phi$.
\end{definition}

For a grammar $G$, we define the {\em derivation} relation $\Rightarrow_G$ on
the sentential forms $\mathcal{SF}(N \cup \Sigma)$ by $\psi_1 \Rightarrow_G \psi_2$
iff there is
some rule $A \to \phi$, such that $\psi_2$ results from $\psi_1$ by
replacing an occurrence of $A$ by $\phi$. As usual,
$\stackrel{*}{\Rightarrow}_G$ denotes the reflexive and transitive closure.
As with Kolam grammars, the derivation thus defined rewrites strings, not pictures.

The derived sentential form denotes a set of pictures. Formally, the picture
language generated by $G$ is the set
\[
L(G) = \{ %\textrm{the smallest picture in }  
p \in \llparenthesis \psi \rrparenthesis \mid \psi \in
\mathcal{SF}(\Sigma), S \stackrel{*}{\Rightarrow}_G \psi \}.
\]
%With a slight abuse of notation, we will often write $A
%\stackrel{*}{\Rightarrow}_G p$, with $A \in N, p \in \Sigma^{*,*}$, instead of
%$\exists \phi : A \stackrel{*}{\Rightarrow}_G \phi, p \in \llparenthesis \phi \rrparenthesis$.

In the literature, parameter $k$ is fixed for a grid grammar $G$, i.e.
all the right parts of rules are either terminal or $k$ by $k$ grids. This
constraint could be relaxed, by allowing different $k$ for different rules:
the results that are shown next still hold for this generalization.

It is trivial to see that grid grammars admit the following normal form:

\begin{definition}\label{def:grid:cnf}
    A grid grammar $G = (\Sigma,N,S,R)$,
    is in {\em Nonterminal Normal Form} (NNF) iff every rule in $R$ has the form
    either $A \to t$, or $A \to [B_{1,1}, \dots, B_{1,k}, \ldots, B_{k,1},$
    $\ldots, B_{k,k}]$, where $A, B_{i,j} \in N$, and $t \in \Sigma$.
\end{definition}

\begin{example}\label{grid-example}

Here is a simple example of a grid grammar in NNF.

\[ 
S \to [S, B, S, B, B, B, S, B, S],
\ \ 
S \to a,
\ \ 
B \to b.
\]

The generated language is that of ``recursive'' crosses of $b$'s in a field
of $a$'s.
Figure \ref{grid:a:picture} shows an example picture of the language.

\begin{figure}[h!]
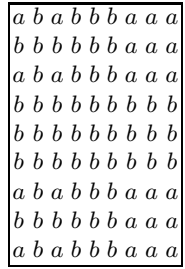

\[
\begin{array}{|ccccccccc|}
\hline
	a & b & a & b & b & b & a & a & a \\
	b & b & b & b & b & b & a & a & a \\
	a & b & a & b & b & b & a & a & a \\
	b & b & b & b & b & b & b & b & b \\
	b & b & b & b & b & b & b & b & b \\
	b & b & b & b & b & b & b & b & b \\
	a & b & a & b & b & b & a & a & a \\
	b & b & b & b & b & b & a & a & a \\
	a & b & a & b & b & b & a & a & a \\
\hline
\end{array}
\]
\caption{A picture of Example \ref{grid-example}.}\label{grid:a:picture}
\end{figure}

\end{example}

\subsubsection{Comparison with other models}

First, we note that this is the only 2D grammatical model presented in this
paper which cannot generate string (i.e. 1D) languages, since all the
generated pictures, if any, have the same number of rows and columns by
definition.

It is easy to see that the class of languages generated by grid grammars
are a proper subset of the one of \prusa{} grammars.  In fact, a grid
grammar can be seen as a particular kind of \prusa{} grammar, in which
symbols in right part of rules generate square pictures having the same
size.

Interestingly, the same construction can be applied also to CF Kolam grammars.

\begin{proposition}\label{grid-in-kolam}
$\LL(GG) \subset \LL(CFKG)$.
%	The class of languages defined by Grid grammars is a proper
%	subset of the class of languages defined by CF Kolam grammars.
\end{proposition}

\begin{pf}
For simplicity, let us consider a grid grammar $G = (\Sigma,N,S,R)$ in NNF.

\begin{itemize}
	\item[(i)] For terminal rules $A \to t, t \in \Sigma$, we introduce the following rules in the
equivalent CF Kolam grammar $G'$:
\[
A \to (A \obar A_v) \ominus (A_h \obar t)
\mid t,\ \
A_h \to A_h \obar t \mid t,
\ \ 
A_v \to t \ominus A_v
\mid t
\]
where $A_h, A_v$ are freshly introduced nonterminals, not used in other rules.
It is easy to see that these rules can only generate all the square
pictures made of
$t$'s.

\item[(ii)] For nonterminal rules $A \to [B_{1,1}, \dots, B_{1,k}, \ldots, B_{k,1}, \ldots,
B_{k,k}]$, we add the following ``structurally equivalent'' kind of rules:
\[
A \to 
\begin{array}{c}
	(B_{k,1} \obar \cdots \obar B_{k,k}) \\
	\ominus \\
	\cdots \\
	\ominus \\
	(B_{1,1} \obar \cdots \obar B_{1,k}) \\
\end{array}
\]
\end{itemize}

To show the equivalence $L(G) = L(G')$, we use induction on derivation
steps. As base case, we note that terminal rules of $G$ are equivalent to
the rules of $G'$ introduced at  (i). 

Induction step: consider a
nonterminal rule like in (ii). By induction hypothesis, all $B_{i,j}$ of
$G'$ generate languages equivalent to their homonym in $G$, and all made of
square pictures. But by definition of $\ominus$, $|(B_{j,1} \obar \cdots
\obar B_{j,k})|_{col}$ $=$  $|(B_{j+1,1} \obar \cdots \obar
B_{j+1,k})|_{col}$,
for all $1 \le j < k$.  Moreover, by definition of $\obar$, $|B_{j,i}|_{row} =
|B_{j,i-1}|_{row}$, for all $1 \le i < k$. Being all squares, this means that
the sentential form $(B_{k,1} \obar \cdots \obar B_{k,k})$ $\ominus \cdots
\ominus$ $(B_{1,1} \obar \cdots \obar B_{1,k})$
of $G'$ generates a picture iff every $B_{i,j}$ have the same size. But
this also means that it is equivalent to the sentential form $[B_{1,1},
\dots, B_{1,k}, \ldots, B_{k,1}, \ldots, B_{k,k}]$ of $G$.   

The inclusion is proper, because by definition grid grammars cannot
generate non-square pictures (e.g. string languages).  \qed
\end{pf}

%%% Local Variables: 
%%% mode: latex
%%% TeX-master: "paper"
%%% End: 

\subsection{Context-free matrix grammars}\label{sec:matrix}

The early model of CF matrix grammars
\cite{GSiromoney-RSiromoney-KKrithivasan:72} is a very
limited kind of CF Kolam grammars. The following definition is taken
and adapted from \cite{MatrixCKY}.

\begin{definition}
%   Let $M = (G,G')$ where $G = (N,T,P,S)$ is a string grammar, where
%   $N$ is the set of nonterminals, $P$ is a set of productions, $S$ is
%   the staring symbol, $T = \{A_1,A_2,\cdots,A_k\}$, $G'$ is a set of
%   string grammars, $G' = \{G_1,G_2,\cdots,G_k\}$ where each $A_i$ is
%   the start symbol of string grammar $G_i$. The grammars in $G'$ are
%   defined over a terminal alphabet $\Sigma$, which is the alphabet of
%   $M$. A grammar $M$ is said to be a {\em Context-Free Matrix Grammar}
%   (CFMG) iff $G$ and all $G_i$ are CF grammars.

  Let $G = (H,V)$ where $H = (\Sigma',N,S,R)$ is a string grammar, where
  $N$ is the set of nonterminals, $R$ is a set of productions, $S$ is
  the starting symbol, $\Sigma' = \{A_1,A_2,\cdots,A_k\}$, $V$ is a set of
  string grammars, $V = \{V_1,V_2,\cdots,V_k\}$ where each $A_i$ is
  the start symbol of string grammar $V_i$. The grammars in $V$ are
  defined over a terminal alphabet $\Sigma$, which is the alphabet of
  $G$. A grammar $G$ is said to be a {\em context-free matrix grammar}
  (CFMG) iff $H$ and all $V_i$ are CF grammars.

Let $p \in \Sigma^{++}$, $p = c_1 \obar c_2 \obar \cdots \obar c_n$. $p \in
L(G)$ iff there exists a string $A_{x_1}A_{x_2}\cdots A_{x_n} \in L(H)$
such that every column $c_j$, seen as a string, is in $L(V_{x_j}), 1 \le j
\le n$. The string $A_{x_1}A_{x_2}\cdots A_{x_n}$ is said to be an {\em
intermediate} string deriving $p$.
\end{definition}

Informally, the grammar $H$ is used to generate a horizontal string of
starting symbols for the ``vertical grammars'' $V_j, 1 \le j \le k$. Then,
the vertical grammars are used to generate the columns of the picture. If
every column has the same height, then the generated picture is defined, and
is in $L(G)$.

\begin{example}\label{matrix:example}
        The language of odd-width
	rectangular pictures over $\{a,b\}$, where the first row, the last
	row, and the central column are made of $b$'s, the rest is filled
	with $a$'s is defined by the CFMG $G_7$ of Figure
        \ref{grammar:matrix}.

\begin{figure}[h!]
\[
\begin{array}{ccl}
G_7 & = & (H,\{V_1,V_2\}) \text{ where } \\
H & : &
S \to A_1 S A_1 \mid A_2 \\
V_1 & : &
A_1 \to b A; \ \ A \to a A \mid b;\\
V_2 & : &
A_2 \to b A_2 \mid b.\\
\end{array}
\]
\medskip
\[p_7 =
\begin{array}{|ccccccc|}
\hline
	b & b & b & b & b & b & b \\
	a & a & a & b & a & a & a \\
	a & a & a & b & a & a & a \\
	a & a & a & b & a & a & a \\
	a & a & a & b & a & a & a \\
	b & b & b & b & b & b & b \\
\hline
\end{array}
\]
\caption{CF matrix grammar $G_7$ of Example \ref{matrix:example}
  (top), 
and an example picture (bottom).
}\label{grammar:matrix}
\end{figure}

\end{example}

\subsubsection{Comparison with other grammar families}\label{matrix:comparison}

First, we note that it is trivial to show that the class of CFMG languages
is a proper subset of CF Kolam languages.

\begin{proposition}
$\LL(CFMG) \subset \LL(CFKG)$.
\end{proposition}

 Intuitively, it is possible to
consider the string sub-grammars $G$, and $G_j$, of a CF matrix grammar
$M$, all in Chomsky Normal Form.
This means that we can define an equivalent $M'$ CF Kolam grammar,
in which rules corresponding to those of $G$ use only the $\obar$ operator,
while rules corresponding to those of $G_j$ use only the $\ominus$
operator.

Also, it is easy to adapt classical string parsing methods to matrix
grammars \cite{MatrixCKY}.

\begin{proposition}
$\LL(CFMG)$ and $\LL(GG)$ are incomparable.
%  The family of CF Matrix grammar languages and the
%  family of Grid grammars languages are incomparable.
\end{proposition}

\begin{pf}
	First, we know that by definition Grid grammars can generate only square pictures.
	On the other hand, it is impossible to define CF matrix grammars
	generating only square pictures. This is because classical string
	pumping lemmata can be applied both to $G$ (the ``horizontal
	component'' of the grammar), and to $G_j, 1 \le j \le k$ (see e.g.
	\cite{NSD89}). Therefore
	the two language classes are incomparable.
	\qed
\end{pf}

%%% Local Variables: 
%%% mode: latex
%%% TeX-master: "paper"
%%% End: 

\section{Summary}\label{SectionConclusion}

We finish with a synopsis of the previous language family
inclusions, and a presentation of the constraints on the tile set of tile
grammars corresponding to each class. 

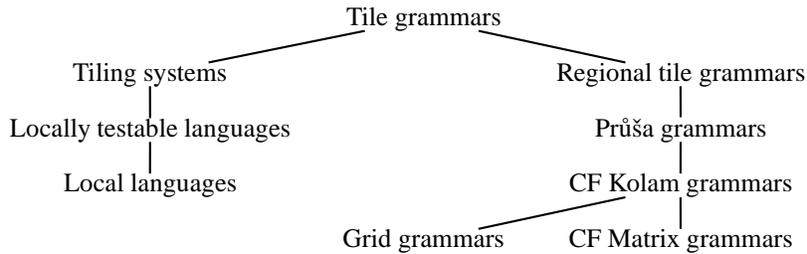
\begin{center}
 \scalebox{1}{%
\psset{arrows=-, colsep=20pt, rowsep=9pt}
\begin{psmatrix}

& \rnode{TRG}{Tile grammars} \\
\rnode{TS}{Tiling systems}& & \rnode{RG}{Regional tile grammars} \\
\rnode{LT}{Locally testable languages}&&\rnode{Prusa}{\prusa{} grammars} \\
\rnode{LOC}{Local languages}&&\rnode{Kolam}{CF Kolam grammars} \\
&  \rnode{Grid}{Grid grammars}&\rnode{Matrix}{CF Matrix grammars}

\ncline{TS}{TRG}
\ncline{RG}{TRG}
%\ncline[linestyle=dashed]{Prusa}{RG}
\ncline{Prusa}{RG}
\ncline{Kolam}{Prusa}
\ncline{Matrix}{Kolam}
\ncline{Grid}{Kolam}

\ncline{TS}{LT}
\ncline{LT}{LOC}
\end{psmatrix}
}
\end{center}

% \begin{center}
%  \scalebox{1}{%
% \psset{arrows=-, colsep=20pt, rowsep=9pt}
% \begin{psmatrix}

%     & \rnode{TRG}{\lng{TG}} \\
%     \rnode{TS}{REC}& & \rnode{RG}{\lng{RTG}} \\
%     &&\rnode{Prusa}{\lng{PG}} \\
%     &&\rnode{Kolam}{\lng{CFKG}} \\
%     \rnode{4DFA}{\lng{4DFA}} &  \rnode{Grid}{\lng{GG}} & \rnode{Matrix}{\lng{CFMG}} \\
%     & \rnode{2RLG}{\lng{2RGL}}  &

% \ncline{TS}{TRG}
% \ncline{RG}{TRG}
% %\ncline[linestyle=dashed]{Prusa}{RG}
% \ncline{Prusa}{RG}
% \ncline{Kolam}{Prusa}
% \ncline{Matrix}{Kolam}
% \ncline{Grid}{Kolam}
% \ncline{TS}{4DFA}
% \ncline{Matrix}{2RLG}
% \ncline{4DFA}{2RLG}

% \end{psmatrix}
% }
% \end{center}

% \subsection{Regional tile grammars}

% Regional tile grammars are tile grammars in which all the nonterminal rules
% $A \to \omega$ are such that $LOC(\omega)$ is a regional language.

\subsection*{\prusa{} grammars}

\prusa{} grammars in NNF are regional tile grammars
with the constraint that tiles used in right part of rules must not
have
one of these forms:
\[
\begin{array}{|cc|}
\hline
A & B \\
C & C \\
\hline
\end{array}
, \ 
\begin{array}{|cc|}
\hline
A & C \\
B & C \\
\hline
\end{array}
, \ 
\begin{array}{|cc|}
\hline
C & C \\
A & B \\
\hline
\end{array}
, \
\begin{array}{|cc|}
\hline
C & A \\
C & B \\
\hline
\end{array}
\]
with $A,B,C$ all different nonterminals. 
(See Proposition \ref{prusa-in-rtg}.)

\subsection*{CF Kolam grammars}

  CF Kolam grammars in CNF can be seen as
  regional tile grammars such that the tile-sets used in the
  right parts of rules must have one of the following forms:
\[
\left\llbracket
\begin{array}{cccccc}
\# & \# & \# & \# & \# & \# \\
\# & A & A & B & B & \# \\
\# & A & A & B & B & \# \\
\# & \# & \# & \# & \# & \# \\
\end{array}
\right\rrbracket,
\ \ 
\left\llbracket
\begin{array}{cccc}
\# & \# & \# & \# \\
\# & A & A &\# \\
\# & A & A & \# \\
\# & B & B &\# \\
\# & B & B & \# \\
\# & \# & \# & \#  \\
\end{array}
\right\rrbracket,
\ \ 
\left\llbracket
\begin{array}{cccc}
\# & \# & \# & \# \\
\# & A & A &\# \\
\# & A & A & \# \\
\# & \# & \# & \#  \\
\end{array}
\right\rrbracket
\]
with $A \ne B$.
(See Proposition \ref{kolam-in-rtg}.)
Clearly, this is also compatible with the constraint of \prusa{} grammars.

\subsection*{Grid grammars}

For grid grammars in NNF, we have the same constraints on nonterminal rules as in
CF Kolam grammars. Moreover, there is a different treatment of terminal
rules of the grid grammar, i.e. rules like $A \to t, t \in \Sigma$.
The corresponding regional tile grammar rules (still maintaining the CF
Kolam grammars constraints) are used to generate from $A$ square
 $t$-homogeneous pictures of any size, and are the following:
%can be obtained by combining the
%constructions used in the proofs of Propositions \ref{kolam-in-rtg}
%and \ref{grid-in-kolam}, namely the following:
\[
A \to
\left\llbracket
\begin{array}{cccc}
\# & \# & \# & \# \\
\# & A_1 & A_1 &\# \\
\# & A_1 & A_1 & \# \\
\# & A_2 & A_2 &\# \\
\# & \# & \# & \#  \\
\end{array}
\right\rrbracket,
\ \ 
A_1 \to
\left\llbracket
\begin{array}{cccccc}
\# & \# & \# & \# & \# \\
\# & A & A & A_3  & \# \\
\# & A & A & A_3  & \# \\
\# & \# & \# & \# & \# \\
\end{array}
\right\rrbracket,
\]
\[
A_2 \to
\left\llbracket
\begin{array}{ccccc}
\# & \# & \# & \# & \# \\
\# & A_4 & A_4 & A_5  & \# \\
\# & \# & \# & \# & \# \\
\end{array}
\right\rrbracket \mid
\left\llbracket
\begin{array}{ccc}
\# & \# & \# \\
\# & A_5  & \# \\
\# &  \# & \# \\
\end{array}
\right\rrbracket, 
\ \ 
A_5 \to t.
\]
\[
A_3 \to
\left\llbracket
\begin{array}{ccc}
\# & \# & \# \\
\# & A_5  & \# \\
\# & A_3  & \# \\
\# & A_3  & \# \\
\# & \# & \# \\
\end{array}
\right\rrbracket \mid
\left\llbracket
\begin{array}{ccc}
\# & \# & \# \\
\# & A_5  & \# \\
\# & \# & \# \\
\end{array}
\right\rrbracket,
\]
with $A_1, \dots, A_5$ all freshly introduced nonterminals.  In
practice, we are using the CF Kolam grammar rules corresponding to
terminal rules of grid grammars of Proposition \ref{grid-in-kolam},
translated into regional tile grammar rules following the construction
of Proposition \ref{kolam-in-rtg}.

\subsection*{CF matrix grammars}

Following the construction sketched in Section
\ref{matrix:comparison} for proving that CF matrix grammars define a subset
of the class defined by CF Kolam grammars, we note that the tile
constraints are exactly the same of CF Kolam grammars. The added constraint
is that if a nonterminal $C$ is used as left part of a ``horizontal'' rule 
\[
C \to
\left\llbracket
\begin{array}{cccccc}
\# & \# & \# & \# & \# & \# \\
\# & A & A & B & B & \# \\
\# & A & A & B & B & \# \\
\# & \# & \# & \# & \# & \# \\
\end{array}
\right\rrbracket
\]
then it shall not be used as left part of a ``vertical'' rule
\[
C \to
\left\llbracket
\begin{array}{cccc}
\# & \# & \# & \# \\
\# & A & A &\# \\
\# & A & A & \# \\
\# & B & B &\# \\
\# & B & B & \# \\
\# & \# & \# & \#  \\
\end{array}
\right\rrbracket
\]
and vice versa.
(This is a direct consequence of the informal considerations at the beginning of 
Section \ref{matrix:comparison} and the proof of Proposition
\ref{kolam-in-rtg}.)

\bigskip

From all that, regional tile grammars prove to be useful as a
unifying, not overly general, concept for hitherto separated grammar
models.

%%% Local Variables: 
%%% mode: latex
%%% TeX-master: "paper"
%%% End: 

\bibliographystyle{plain}
\bibliography{2Dbib}

\begin{thebibliography}{10}

\bibitem{Wang}
C.~Allauzen and B.~Durand.
\newblock Tiling problems.
\newblock In E.~Borger and E.~Gradel, editors, {\em The classical decision
  problem}. Springer-Verlag, 1997.

\bibitem{RTG1}
Alessandra Cherubini, Stefano {Crespi Reghizzi}, and Matteo Pradella.
\newblock Regional languages and tiling: A unifying approach to picture
  grammars.
\newblock In {\em Mathematical Foundations of Computer Science (MFCS 2008)},
  volume 5162 of {\em Lecture Notes in Computer Science}, pages 253--264.
  Springer, 2008.

\bibitem{CCPS}
Alessandra Cherubini, Stefano {Crespi Reghizzi}, Matteo Pradella, and Pierluigi
  {San Pietro}.
\newblock Picture languages: Tiling systems versus tile rewriting grammars.
\newblock {\em Theoretical Computer Science}, 356(1-2):90--103, 2006.

\bibitem{TRG1}
S.~{Crespi Reghizzi} and M.~Pradella.
\newblock {T}ile {R}ewriting {G}rammars and {P}icture {L}anguages.
\newblock {\em Theoretical Computer Science}, 340(2):257--272, 2005.

\bibitem{KolamParsing}
S.~{Crespi Reghizzi} and M.~Pradella.
\newblock A {CKY} parser for picture grammars.
\newblock {\em Information Processing Letters}, 105(6):213--217, 2008.

\bibitem{labelled-wang1997}
Lucio de~Prophetis and Stefano Varricchio.
\newblock Recognizability of rectangular pictures by {W}ang systems.
\newblock {\em Journal of Automata, Languages and Combinatorics},
  2(4):269--288, 1997.

\bibitem{Drewes1996}
Frank Drewes.
\newblock Language theoretic and algorithmic properties of d-dimensional
  collages and patterns in a grid.
\newblock {\em Journal of Computer and System Sciences}, 53(1):33--66, 1996.

\bibitem{Drewes2003}
Frank Drewes, Sigrid Ewert, Renate Klempien-Hinrichs, and Hans-J{\"o}rg
  Kreowski.
\newblock Computing raster images from grid picture grammars.
\newblock {\em Journal of Automata, Languages and Combinatorics},
  8(3):499--519, 2003.

\bibitem{finkel1974}
Raphael~A. Finkel and Jon~Louis Bentley.
\newblock Quad trees: A data structure for retrieval on composite keys.
\newblock {\em Acta Informatica}, 4:1--9, 1974.

\bibitem{GR-recognizable-pl}
D.~Giammarresi and A.~Restivo.
\newblock Recognizable picture languages.
\newblock {\em International Journal Pattern Recognition and Artificial
  Intelligence}, 6(2-3):241--256, 1992.
\newblock Special Issue on {\em Parallel Image Processing}.

\bibitem{GR-two-dl}
D.~Giammarresi and A.~Restivo.
\newblock Two-dimensional languages.
\newblock In Arto Salomaa and Grzegorz Rozenberg, editors, {\em Handbook of
  Formal Languages}, volume 3, Beyond Words, pages 215--267. Springer-Verlag,
  Berlin, 1997.

\bibitem{Harrison78}
M.~A. Harrison.
\newblock {\em Introduction to Formal Language Theory}.
\newblock Addison Wesley, 1978.

\bibitem{lewis}
H.~Lewis.
\newblock Complexity of solvable cases of the decision problem for predicate
  calculus.
\newblock In {\em Proc. 19th {S}ymposium on {F}oundations of {C}omputer
  {S}cience}, pages 35--47, 1978.

\bibitem{lindgren}
K.~Lindgren, C.~Moore, and M.~Nordahl.
\newblock Complexity of two-dimensional patterns.
\newblock {\em Journal of Statistical Physics}, 91(5-6):909--951, June 1998.

\bibitem{STACS::Matz1997}
O.~Matz.
\newblock Regular expressions and context-free grammars for picture languages.
\newblock In {\em 14th Annual Symposium on Theoretical Aspects of Computer
  Science}, volume 1200 of {\em Lecture Notes in Computer Science}, pages
  283--294, 1997.

\bibitem{NSD89}
M.~Nivat, A.~Saoudi, and V.~R. Dare.
\newblock Parallel generation of finite images.
\newblock {\em International Journal Pattern Recognition and Artificial
  Intelligence}, 3(3-4):279--294, 1989.

\bibitem{Prusa2001}
D.~Pr{\r{u}}{\v s}a.
\newblock Two-dimensional context-free grammars.
\newblock In G.~Andrejkova and S.~Krajci, editors, {\em Proceedings of ITAT
  2001}, pages 27--40, 2001.

\bibitem{Prusa2004}
D.~Pr{\r{u}}{\v s}a.
\newblock {\em Two-dimensional Languages ({PhD Thesis})}.
\newblock Charles University, Faculty of Mathematics and Physics, Czech
  Republic, 2004.

\bibitem{MatrixCKY}
V.~Radhakrishnan, V.~T. Chakaravarthy, and K.~Krithivasan.
\newblock Pattern matching in matrix grammars.
\newblock {\em Journal of Automata, Languages and Combinatorics}, 3(1):59--72,
  1998.

\bibitem{Simplot99}
David Simplot.
\newblock A characterization of recognizable picture languages by tilings by
  finite sets.
\newblock {\em Theoretical Computer Science}, 218:297--323, 1999.

\bibitem{GSiromoney-RSiromoney-KKrithivasan:72}
G.~Siromoney, R.~Siromoney, and K.~Krithivasan.
\newblock Abstract families of matrices and picture languages.
\newblock {\em Computer Graphics and Image Processing}, 1:284--307, 1972.

\bibitem{GSiromoney-RSiromoney-KKrithivasan:73b}
G.~Siromoney, R.~Siromoney, and K.~Krithivasan.
\newblock Picture languages with array rewriting rules.
\newblock {\em Information and Control}, 23(5):447--470, 1973.

\bibitem{CKY}
D.~H. Younger.
\newblock Recognition of context-free languages in time $n^3$.
\newblock {\em Information and Control}, 10(2):189--208, 1967.

\end{thebibliography}

\end{document}